\UseRawInputEncoding
\documentclass[aps,prb,twocolumn,showpacs,preprintnumbers,amsmath,amssymb,superscriptaddress,floatfix]{revtex4-2}
\usepackage{graphicx}
\usepackage{hyperref}
\usepackage{amsfonts}
\usepackage{CJK}
\usepackage{float}
\begin{document}

%\title{Characterizing Topological Properties of one-dimensional non-Hermitian
%  interacting systems}
\title{Characterizing the Bulk-Boundary Correspondence of one-dimensional non-Hermitian
    interacting systems by edge entanglement entropy}
\author{Weitao Chen}
%\affiliation{Department of Physics, Xiamen University, Xiamen 361005, Fujian, China}
\author{Liangtao Peng} 
\affiliation{Department of Physics, Xiamen University, Xiamen 361005, China} 
\author{Hantao Lu}\email{luht@lzu.edu.cn}
\affiliation{School of Physical Science and Technology $\&$ Key
  Laboratory for Magnetism and Magnetic Materials of the MoE, Lanzhou
  University, Lanzhou 730000, China} 
\author{Xiancong Lu}\email{xlu@xmu.edu.cn} 
\affiliation{Department of Physics, Xiamen University, Xiamen 361005, China}

\date{\today}
\begin{abstract}
  Dramatically different from the Hermitian systems, the conventional
  Bulk-Boundary Correspondence (BBC) is broken in the non-Hermitian
  systems. In this article, we use edge entanglement entropy to
  characterize the topological properties of non-Hermitian
  Su-Schrieffer-Heeger Hubbard model.
%We show that the entanglement entropy shows scaling behavior at the topological
%transition points, as well as the gap close points, and therefore is
%a good indicator for the breakdown of BBC.
  For free Fermions, we study the scaling behavior of entanglement
  entropy and demonstrate that the edge entanglement entropy is a good
  indicator to delimit different phases of non-Hermitian systems.  We
  further generalize the edge entanglement entropy to the 
  non-Hermitian interacting Hubbard chain, and obtain the topological phase
  diagram in the plane of interaction and non-Hermitian hopping
  amplitudes. It is found that the Hubbard interaction diminishes and
  weakens the breakdown of Bulk-Boundary Correspondence, which
  eventually disappears at some critical value of interaction.
\end{abstract}

\pacs{PACS}
%\keywords{}
\maketitle

%===============================================================================
\section{Introduction}\label{sec1}

Fascinating phenomena emerge in non-Hermitian systems with gain and
loss of energy
\cite{ca.wi.15,di.ma.16,ka.as.17,li.ge.12,oz.pr.19,fl.so.15,do.ma.16,ma.po.15,ch.ji.14,xu.ma.16,ga.es.15,ma.el.08,
  gu.sa.09,ru.ma.10,li.Ra.11,le.ch.14}, which attract considerable
attention from many fields of physics \cite{as.go.20}. Various
non-Hermitian systems have been intensively investigated in recent
years, such as acoustics\cite{ma.sh.16,cu.ch.16}, optics
\cite{za.fl.19,el.ma.18,mi.al.19}, ultra-cold atoms
\cite{dale.14,kuhr.16,li.ha.19,la.ja.19,re.li.21}, parity-time-symmetric systems
\cite{be.07,do.du.07,oz.ro.19}, driven-dissipative systems
\cite{de.ha.10,be.li.06,ri.do.13,si.bu.16,we.ks.21}, and material
junctions \cite{be.bu.19,ca.bl.21,sa.ca.16,pi.na.13,pi.na.12}.  From the
theoretical point of view, significant progress has been made on
understanding the topological properties of non-Hermitian systems, by
generalizing the conceptions from Hermitian topological band theory
\cite{qi.zh.11,ha.ka.10}. Especially, the bulk-boundary correspondence
(BBC), which states that topological invariants computed from the
Bloch Hamiltonian under periodic boundary condition (PBC) exactly
correspond to the boundary states under open boundary condition (OBC)
\cite{ry.ha.06,gr.ho.13,rh.be.17}, is a general principle of
topological theory for Hermitian systems. However, it is shown that
the BBC is broken in the non-Hermitian system
\cite{lee.16,xion.18,ya.wa.18,ku.ed.18}, due to the sensitivity of
energy spectrum to the boundary conditions.  The mechanism of the
breakdown of BBC has been studied
\cite{be.bu.21,ya.wa.18,ku.ed.18,ok.ka.20,zh.ya.20,bo.kr.20,ji.so.19,zi.re.21}
and methods to reconstruct the BBC have been proposed, \textit{e.g.},
the generalized Brillouin zone (GBZ)
\cite{ya.wa.18,yo.mu.19,le.th.19,im.ta.19,ok.ka.20,zh.ya.20,ya.zh.20},
biorthogonal polarization \cite{ku.ed.18}, and singular value
decomposition \cite{he.ba.19}.

% By determining distinct band topology,
%non-Hermitian gapped phases are classified into two category of band
%stuctures,i.e., point gaps and line gaps \cite{ka.sh.19}.

% An exotic effect named
%5non-Hermitian skin effect was firstly reported in
%ref. \cite{ya.wa.18},  

%\cite{bo.kr.20,lo.19,al.va.18,le.th.19,so.ya.19,ji.so.19,le.li.19,zh.ya.20,ok.ka.20,ho.he.20}

The non-Hermitian systems on many-body basis show different behaviors
comparing with that on single-particle level. For example, the
well-established non-Hermitian skin effect \cite{ya.wa.18} does not
appear within the many-body framework, due to the Pauli exclusion
principle \cite{mu.le.20,le.le.20}. For interacting many-body systems,
quite a lot of novel non-Hermitian phenomena have been revealed, such
as the many-body localization \cite{ha.ka.19}, the erosion of real-space
Fermi surface \cite{mu.le.20}, non-Hermitian topological Mott phases
in both bosonic and fermionic superlattices
\cite{zh.ch.20,xu.ch.20,li.he.20}, the emergence of pseudospectrum in
correlated Hermitian systems \cite{ok.sa.21b,yosh.21}, the
classification of topological phases in one and zero dimension
\cite{xi.zh.21,yo.ha.21}, off-diagonal long-range order with attractive
interaction \cite{zh.so.21}, and skin clusters from strong interactions
\cite{sh.le.21}.

As far as the many-body physics is concerned, quantum entanglement
provides a unique perspective and has been proved to be a powerful
tool to understand the phases of Hermitian systems
\cite{ki.pr.06,le.we.06,lafl.16}.
%Apart from that, the concept of quantum entanglement had contributed a
%novel perspective to study the quantum many-body system
%\cite{ki.pr.06,le.we.06,lafl.16}. 
For instance, the entanglement
entropy (EE) was employed to identify topological orders in long-range
entangled systems like quantum spin liquid
\cite{zh.gr.11,zh.gr.12,ya.hu.11,ji.wa.12,ji.ya.12}.
%For short-range
%entangled systems, edge entanglement entropy was proposed to determine
%the edge degeneracy of Hermitian topological systems.
For the one-dimensional (1D) gapped systems,  
the entanglement entropy of ground state obeys the well-known area-law
\cite{ha.ma.01,hast.07,wo.ve.08}, whereas it exhibits logarithmic
scaling behavior for a gapless critical chain, with the emergence of
conformal field theory (CFT) \cite{ho.la.94,ca.ca.09,ch.yo.20}.
A natural question is what is the entanglement properties of 
non-Hermitian systems. Recently, this topic receives significant
progress, \textit{e.g.}, the conception of EE has been successfully extended to the
non-Hermitian many-body systems
\cite{co.ja.17,he.re.19,ch.yo.20,le.le.20,mu.le.20,ch.ch.21,ok.sa.21a,ba.do.21,mo.ma.21,gu.yu.21,tu.tz.21}.
In particular, the scaling behavior of EE at critical points as well
as the normal phases have been well studied for various non-Hermitian
non-interacting models
\cite{he.re.19,ch.yo.20,mu.le.20,gu.yu.21,tu.tz.21,lee.20}. 
%Whereas, the studies of
%entanglement properties for interacting non-Hermitian systems are
%still at the beginnings \cite{mu.le.20}. 
%However, very few works investigate the entanglement properties of
%interacting non-Hermitian many-body model, and moreover how the
%entanglment behaiver is modified by interaction and its relation to
%topology have not been sufficiently explored.
In spite of these works on free Fermions, the entanglement properties
of interacting non-Hermitian systems have not been fully explored yet,
and how the entanglement
behavior is modified by interaction as well as its relation to
topology remain unclear.

In this paper, we address this issue by studying the non-Hermitian
Su-Schrieffer-Heeger (SSH) model with Hubbard interaction (SSHH), using the Exact
Diagonalization (ED) method. We will specifically focus on the edge
entanglement entropy, which is firstly introduced to measure edge
degeneracy of topological states in Hermitian SSHH model
\cite{wa.xu.15}.
%For the interacting Hermitian Su-Schrieffer-Heeger (SSH) model, a
%quantity termed edge entanglement entropy is introduced to measure
%edge degeneracy of topological states \cite{wa.xu.15}. 
It is proposed in Ref. \cite{le.le.20} that the edge EE is a useful
tool to detect the topological properties of non-Hermitian many-body
systems. However, only non-interacting Hamiltonian is tested there,
and interacting non-Hermitian models have not been verified yet, which
is our main starting point.  We will show in this paper that the edge
EE is a good indicator of the breakdown of BBC for non-Hermitan
systems. Based on edge EE, we demonstrate that the breakdown of
BBC is diminished as the Hubbard interaction is increasing.
%interplay with the non-Hermicity will be demonstrated.
The paper is organized as follows: In
Sec.~\ref{sec2}, we introduce the Hamiltonian of non-Hermitian model
and review the methods to calculate the EE. In Sec.~\ref{sec3}, we
present the results of EE in a non-interacting non-Hermitian SSH
model, in which the scaling behavior and phase transitions are
studied. In Sec.~\ref{sec4}, we present the phase diagram for
interacting non-Hermitian model, according to the edge entanglement
entropy. A brief summary is given in Sec.~\ref{sec5}.

%extending the notion of topological phases to NH systems has become a
%broad frontier of current research.

%===============================================================================
\section{Model and entanglement entropy}\label{sec2}

\begin{figure}[t]
    \centering
    \includegraphics[width=0.98\linewidth]{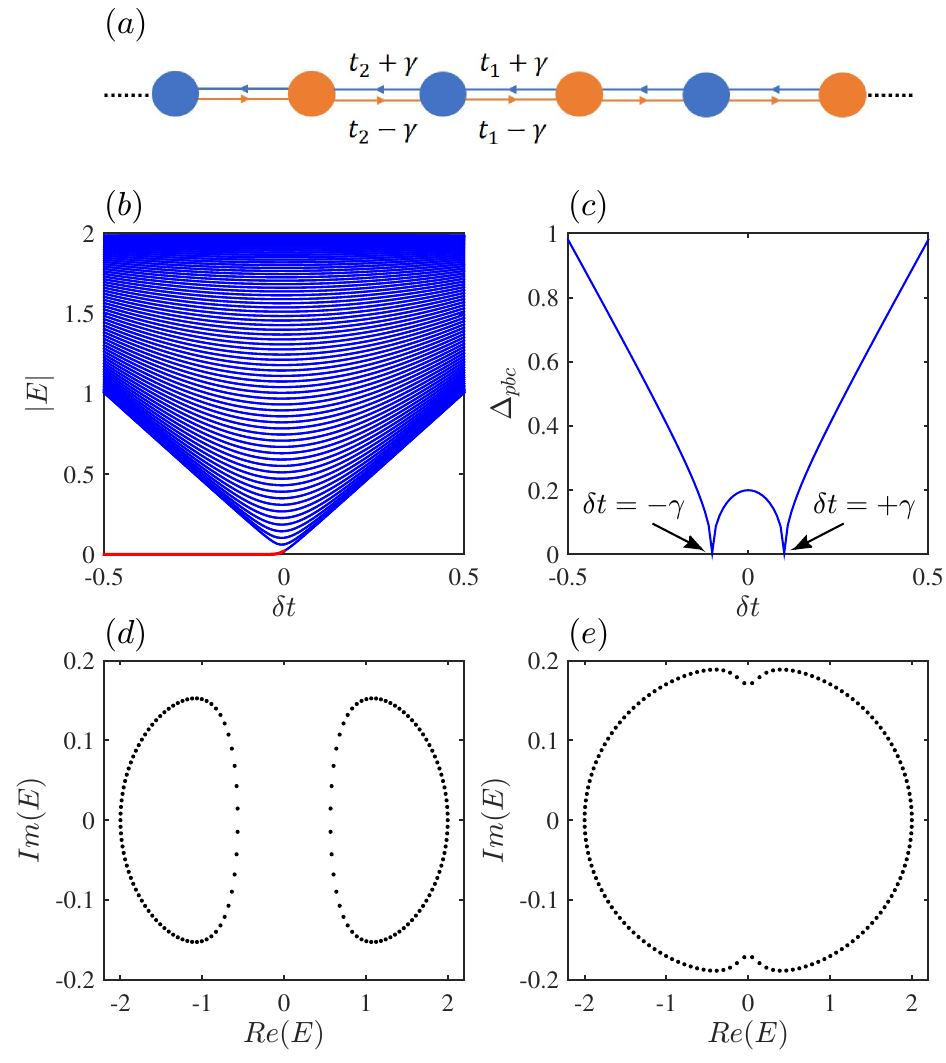}
    \caption{(a) Schematic picture for the non-Hermitian SSHH model
      (\ref{model}). (b) The energy spectrum $|E|$ under OBC for the
      non-interacting fermions with the number of lattice sizes
      $L=150$. The red line denotes the zero modes. 
      (c) The gap $\Delta_{pbc}$ extracted from the absolute
      values of Bloch band with PBC. Two types of energy spectrum
      under PBC, showing the line gap (d) and point gap (e) with
      $\delta t=-0.3$ and $\delta t=-0.05$, respectively. The
      values of other parameters are $U=0$ and $\gamma=0.1$. }
    \label{fig1}
\end{figure}

We consider the 1D spinful non-Hermitian version of SSHH model,
described by the following Hamiltonian
\begin{eqnarray}\label{model}
H = &-&\sum_{i,\sigma }\Big\{ [t-(-1)^{i}\delta t ]
               c_{i\sigma }^{\dagger}c_{i+1,\sigma } + H.c. \Big\} \nonumber\\
    &+&\gamma \sum_{i,\sigma } ( c_{i\sigma }^{\dagger }c_{i+1,\sigma}
          - c_{i+1,\sigma }^{\dagger }c_{i\sigma } )
          + U \sum_{i}n_{i\uparrow}n_{i\downarrow}
\end{eqnarray}
where $c_{i\sigma }^{\dagger }$ ($c_{i\sigma }$) is the fermionic
creation (annihilation) operator at the $i$-th site with spin $\sigma$
($\sigma = \uparrow,\downarrow$), 
%$t$ is the hopping amplitude and
%$\delta t$ is the dimerization parameter, 
$t_1=-(t+\delta t)$ and $t_2=-(t-\delta t)$ are the hopping amplitudes
inside and between unit cells, $\gamma$ denotes the non-reciprocal
contribution to the hopping,
$n_{i\sigma}=c_{i\sigma }^{\dagger }c_{i\sigma }$ is particle number,
and $U$ is the on-site Hubbard interaction. 
In the remainder of this paper, we set $t$ as the
energy unit of the system, \textit{i.e.}, $t=1$.
The second term in Hamiltonian (\ref{model}) with $\gamma\neq 0$ is the source of
non-Hermicity, in which the part hopping to the left is different from
that hopping to the right (its
Hermitian conjugate) leading to the directional localization of bulk states.
This model, with a
schematic figure shown in Fig. \ref{fig1}(a), can be experimentally
realizable in an ultracold fermionic system with atom loss
\cite{li.ha.19,la.ja.19,re.li.21}. 
The interaction can be easily tuned through Feshbach resonances, the
hopping amplitude is controlled by the depth of optical lattices,
while the non-reciprocal term can be realized by laser-induced atom
loss \cite{li.ha.19,li.le.20}.

%In our research, most computation was done in many-body basis states
%or Fock basis states (Except for the computation of single-particle
%energy spectrum). In these basis states, the Pauli exclusion principle
%is automatically satisfied.

To investigate the non-Hermitian model (\ref{model}), we use the
biorthogonal formulation of quantum mechanics \cite{brod.13}. For a
diagonalizable non-Hermitian Hamiltonian $H$, one have
\begin{equation}
H\left | \Phi_{R,n} \right \rangle = E_{n} \left | \Phi_{R,n} \right \rangle,\quad
H^{\dagger}\left | \Phi_{L,n} \right \rangle = E_{n}^{\ast}\left | \Phi_{L,n} \right \rangle.
\end{equation}
Here $\left | \Phi_{L,n} \right \rangle$ and
$\left | \Phi_{R,n} \right \rangle$ are left and right eigenvectors,
which can be chosen to satisfy the biorthonormality condition
$\left \langle \Phi_{L,m} | \Phi_{R,n} \right \rangle =
\delta_{mn}$. The biorthonormal expectation value of an observable $\hat{A}$ can be
computed using both left and right states of a system,
$\langle\hat{A}\rangle = \langle \Psi_{L} |\hat{A}|\Psi_{R}\rangle$,
in which any right state $|\Psi_{R}\rangle$ can be decomposed into the eigenstates
$|\Psi_{R}\rangle=\sum_n C_n |\Phi_{R,n}\rangle$ and the corresponding
left state is defined as
$|\Psi_{L}\rangle=\sum_n C_n |\Phi_{L,n}\rangle$ such that $\langle \Psi_{L}|\Psi_{R}\rangle=1$
\cite{le.le.20,he.re.19}. 
%Many-body
%eigenstates can be constructed by ''filling'' up the energy levels $E_{n}$
%of the Hamiltonian \cite{le.le.20,ch.yo.20}. 
For non-Hermitian systems with complex energy $E_{n}$, there is no
longer a natural way to define the many-body ground state, and
different schemes of definition have been adopted before
\cite{ch.yo.20,he.re.19,gu.yu.21}. In this paper, we choose the left
and right many-body ground states to be the ones that have the lowest
real part energy, \textit{i.e.}, by filling up the levels to Fermi
energy according to the real part of the energy, as has been used in Refs.
\cite{le.le.20,ch.yo.20,he.re.19,gu.yu.21}.

Entanglement is one of the most fundamental properties of many-body
quantum state, which is measured by EE, a quantity
built from density matrix.
Based on the biorthogonal formulation, the density matrix of
non-Hermitian system is defined as
$\rho^{RL}=|\Psi_{R}\rangle\langle \Psi_{L} |$, which in general is 
non-Hermitian $(\rho^{RL})^\dagger\neq
\rho^{RL}$. 
Note that another type of density matrix is given by
$\rho^{RR}=|\Psi_{R}\rangle\langle \Psi_{R} |$, which also contains
information of the system \cite{he.re.19,mo.ma.21} but is not studied in this paper.
By partitioning the total
system into two subsystems $A$ and $B$, and then taking the partial
trace over the subsystem $B$, the reduced density matrix of subsystem
$A$ can be calculated $\rho^{RL}_A=Tr_B \rho^{RL}$. The entanglement entropy is
the von-Neumann entropy of non-Hermitian reduced density matrix,
\begin{equation}\label{S}
S\equiv -Tr\left ( \rho^{RL}_{A}\ln \rho^{RL}_{A} \right ),
\end{equation}
which is a straightforward generalization of the definition for
Hermitian system \cite{he.re.19,le.le.20,ch.yo.20}. Note that the
non-Hermitian entanglement entropy given in Eq. (\ref{S}) can be negative or
complex \cite{he.re.19,ch.yo.20,lee.20,tu.tz.21},
due to the possible negative or complex eigenvalues of $\rho^{RL}_A$. This
challenges the probability interpretation of the eigenvalues
of density matrix \cite{tu.tz.21}.  
The higher order
($\alpha$ order) entanglement entropy, Renyi entanglement entropy, for
non-Hermitian systems is defined as
\begin{equation}\label{eq2}
S_{\alpha }\equiv \frac{1}{1-\alpha }\ln Tr\left [ \left (\rho^{RL}_{A} \right )^{\alpha } \right ].
\end{equation}
The von-Neumann entanglement entropy can be obtained from Renyi
entropy by letting $\alpha \rightarrow 1$:
$S\equiv S_1 = \lim_{\alpha\rightarrow 1}S_{\alpha}$.
Based on the Renyi entanglement entropy, the edge entanglement entropy
is given by \cite{wa.xu.15,le.le.20}
\begin{equation}\label{eq:Sedge}
S_{\alpha ,edge}\equiv S_{\alpha ,OBC} - \frac{1}{2} S_{\alpha ,PBC},
\end{equation}
where $S_{\alpha ,OBC}$ and $S_{\alpha ,PBC}$ are calculated under OBC
and PBC, respectively. Due to the area law that entanglement entropy
obeys, the difference between $S_{\alpha,OBC}$ and half of
$S_{\alpha,PBC}$ contains the topological information of edge states
such as the degeneracy \cite{wa.xu.15}. Especially for the Hermtian
SSH model, the $S_{2,edge}$ exhibits a quantized value $2\ln 2$ ($0$)
in the gapped topological (trivial) phase \cite{ry.ha.06,wa.xu.15}.
For the non-Hermitian non-interacting SSH model, a similar quantized
behavior was observed \cite{le.le.20}, but only in a specific
regions as will be shown in the next section.

For the free Fermions, the correlation matrix is an efficient technique
to compute the reduced density matrix and the entanglement entropy
\cite{pesc.03,ch.he.04,pe.ei.09}. It is shown that, within the
framework of biorthogonal formulation, the method of correlated matrix
is also valid for the non-interacting non-Hermitian system
\cite{he.re.19,ch.yo.20}. The key procedure is to write the reduced
density matrix as $\rho^{RL}_A=\frac{1}{Z}\exp{(-\widetilde{H})}$,
with
$\widetilde H = \sum_{\alpha\beta} h_{\alpha\beta} c_\alpha^\dagger
c_\beta$ being the entanglement Hamiltonian. The Hamiltonian matrix
$\mathbf h$ can be written in terms of the two-site correlation matrix
$\mathbf{C}^A$ \cite{he.re.19,ch.yo.20}, \textit{i.e.},
$\mathbf{h}=\ln [(1-\mathbf{C}^A)/\mathbf{C}^A]$, where the elements
of $\mathbf{C}^A$ are restricted to the subsystem $A$ and are defined
as
$C^A_{ij}=\langle \Psi_L | c^\dagger_i c_j | \Psi_R\rangle = Tr (
\rho_A^{RL} c^\dagger_i c_j)$. Once the eigenvalues $\xi_l$ of matrix
$\mathbf{C}^A$ is known by diagonalization, the entanglement spectrum
of $\rho_A^{RL}$ is given by $\epsilon_l=\ln [(1-\xi_l)/\xi_l]$, and
thus the entanglement entropy as well
\begin{eqnarray}
S &=& -\sum_l \Big[ \xi_l\ln\xi_l + (1-\xi_l)\ln (1-\xi_l) \Big]\\
S_\alpha &=& \frac{1}{1-\alpha} \sum_l 
           \ln \Big[ (1-\xi_l)^\alpha + \xi^\alpha_l \Big].
\end{eqnarray}

For the interacting non-Hermitian system, the method of correlated
matrix is no longer valid. We then turn to the non-Hermitian Lanczos
ED method \cite{ba.de.00} to compute the entanglement entropy. 
In practice, the ED method begins with representing the many-body
basis in Hilbert space. For the model in Eq. (\ref{model}), a
convenient basis can be constructed in real space as
$|\psi\rangle=\prod_{i=1}^L
(c_{i\uparrow}^\dagger)^{n_i^\uparrow}\prod_{j=1}^L
(c_{j\downarrow}^\dagger)^{n_j^\downarrow}|0\rangle$, where
$n_i^\sigma\in\{0,1\}$ indicates whether or not site $i$ is occupied
by a spin-$\sigma$ electron. A sequence of values $\{ n_i^\sigma \}$
can be interpreted as a bit-pattern, which uniquely corresponds to an
integer $I^\sigma$. In this way, each basis-state is represented by a
pair of integers $(I^\uparrow, I^\downarrow)$
\cite{lin.90,li.gu.93}. After generating the basis states in order,
the elements of Hamiltonian matrix can be computed by suitable
bit-level operations \cite{li.gu.93}.  The Hamiltonian matrix is
sparse in this representation, and only nonzero elements are stored to
save memory and speed up operations. The following key step is to
diagonalize the Hamiltonian matrix using non-Hermitian Lanczos method,
which is a two-sided iterative algorithm \cite{ba.de.00}.
By using the relation of recurrence, two
sequences of Lanczos vectors can be generated, which are biorthogonal
and span the left and right Krylov spaces. The Hamilton matrix in these
Lanczos bases is a non-Hermitian tridiagonal matrix whose eigenvectors
can be used to construct the eigenvectors of the original Hamiltonian
\cite{lin.90,li.gu.93,ba.de.00}. The right (left) ground state vector obtained from
Lanczos can be projected onto a right (left) tensor with a rank $L$,
the number of total lattice sites. After cutting the system, the right
(left) tensor can be further reshaped into a $4^{L_A}$-by-$4^{L_B}$
right (left) matrix $T_R$ ($T_L$), where $L_A$ and $L_B$ are the
lattice sizes of subsystems $A$ and $B$, respectively. The matrix
$T_R$ ($T_L$) corresponds to a direct product of the right (left)
eigenvectors of $A$ and $B$. The partial trace over $B$ can be easily
performed by matrix multiplication and the reduced density matrix can
be calculated by $\rho^{RL}_A=T_R^\dagger T_L$.
With $\rho^{RL}_A$, the entanglement entropy can be worked out
through their definitions in Eqs. (\ref{S}) and (\ref{eq2}).

\begin{figure}[t]
    \centering
    \includegraphics[width=0.98\linewidth]{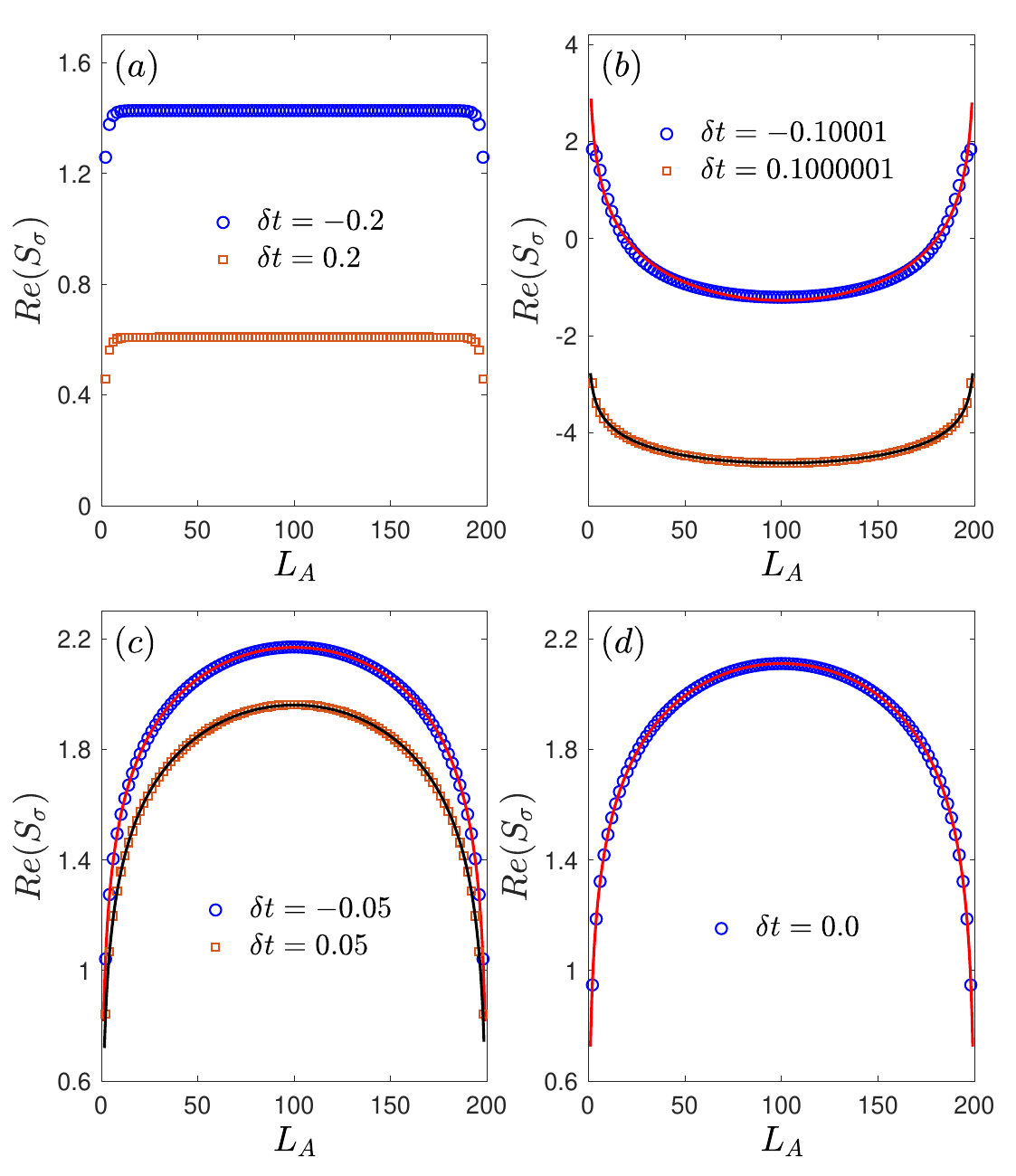}
    \caption{The scaling behavior of the real part of von-Neumann
      entanglement entropy $S_\sigma$ for one spin component
      ($\sigma=\uparrow,\downarrow$) of non-Hermitian free
      fermions. The system is under PBC and at half-filling. The
      values of hopping dimerization $\delta t$ in each sub-figures
      are: (a) $\delta t=-0.2$ and 0.2; (b) $\delta t=-0.10001$ and 0.1000001;
      (c) $\delta t=-0.05$ and 0.05; (d) $\delta t=0.0$. The values of
      other parameters are $U=0$, $\gamma=0.1$, and total
      lattice sizes $L=200$. The solid lines in sub-figures (b), (c)
      and (d) are the fitting curves using Eq. (\ref{scaling}) with
      $\alpha=1$.}
    \label{fig2}
\end{figure}

\section{Non-Hermitian Free-Fermionic SSH model}\label{sec3}

We firstly focus on the non-interacting system with $U=0$. In this
case, the non-Hermitian SSH model shows the breakdown of conventional
BBC \cite{lee.16,xion.18,ya.wa.18,ku.ed.18}, \textit{i.e.}, the
topological transition points extracted from the edge states of OBC
differ from the gap-closing points of the Bloch Hamiltonian under
PBC. For the model in Eq. (\ref{model}), the topological transition
point locates at $\delta t=0$, whereas the gap-closing points are at
$\delta t = \pm \gamma$, as shown in Fig. \ref{fig1}(b) and (c) where
the energy spectrum of OBC and absolute gap of PBC are plotted. These
three transition points delimit four different phases in the whole
parameter region. Moreover, two types of gaps can be observed: the
point gap within $-\gamma<\delta t<\gamma$ and the line gap when
$\delta t>\gamma$ and $\delta t<-\gamma$; see Fig. \ref{fig1}(d) and
(e) for typical examples. The type of gap will greatly influence the
topological properties of the Hamiltonian \cite{ka.sh.19,he.re.19,be.bu.21}.

%Since only state with real part energy survives at a large time
%scale,
 
The real part of
von-Neumann entanglement entropy under PBC is plotted in
Fig. \ref{fig2} against the subsystem size $L_A$, with a fixed total
lattice size $L=200$. To study the critical behavior, we fit the
entanglement entropy using the following universal formula
\cite{ca.ca.04,lafl.16}
\begin{eqnarray}\label{scaling}
S_\alpha(L_A)=\frac{c}{6} \Big(1+\frac{1}{\alpha}\Big)
    \ln \Big( \frac{L}{\pi} \sin \Big[ \frac{\pi L_A}{L}\Big] \Big)
    +s_\alpha+\cdots,
\end{eqnarray}
where $c$ is the central charge described by conformal field theory
(CFT), $s_\alpha$ is a non-universal constant term, and $\alpha$ is
the order of Renyi entropy which is chosen to be
$\alpha=1$ for the case of von-Neumann entropy.
As will be shown below, the non-Hermitian entanglement entropy
exhibits scaling behavior of Eq. (\ref{scaling}) around the critical
points as well as in the point gap phases.

\begin{figure}[t]
    \centering
    \includegraphics[width=0.95\linewidth]{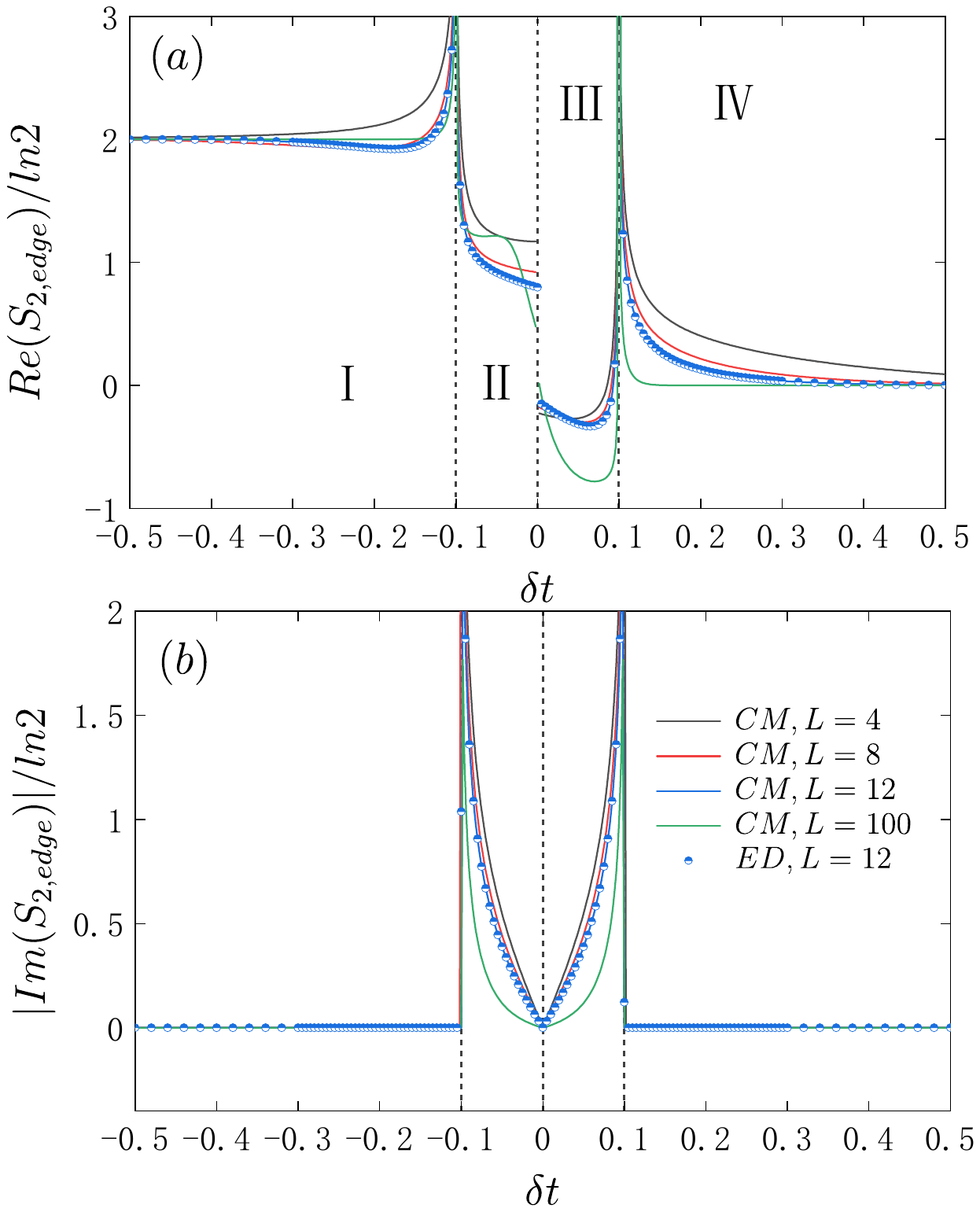}
    \caption{The second order edge entanglement entropy $S_{2 ,edge}$,
      calculated by the correlation matrix (CM) method (lines, with $L=4$,
      $8$, $12$, and
      $100$) and ED method (blue circle, $L=12$), as a function of
      the hopping dimerization $\delta t$. The system is at
      half-filling with $N_{\uparrow}=N_{\downarrow}=L/2$,
      $U=0$, and $\gamma=0.1$. Four phases, from $I$ to $IV$, can be
      discriminated by using two gap-closing points at
      $\delta t=\pm \gamma$ and one critical point at $\delta t=0$.}
    \label{fig3}
\end{figure}

In the regions $\delta t>\gamma$ and $\delta t<-\gamma$, the
von-Neumann EE is always real and gets saturated to a constant value
as the increase of subsystem size $L_A$, implying that the system is
non-critical and is short-range correlated; See Fig. \ref{fig2}(a) for
typical examples. This behavior is related to the gap structure of
energy band, \textit{i.e.}, the real line gap as shown in
Fig. \ref{fig1}(d). In this case, non-Hermitian Hamiltonian may be
continuously deformed into a Hermitian Hamiltonian without breaking of
symmetries \cite{he.re.19}. At $\delta t=\pm\gamma$, the line gap
transforms into the point gap (gap-closing points), and therefore a
PBC phase transition occurs. The real part of von-Neumman EE, shown in
Fig. \ref{fig2}(b), displays a logarithmic dependence on the subsystem
size $L_A$, when shifting away from the critical points $\delta t=\pm 0.1$ by a tiny
value $\epsilon$ \cite{ch.yo.20,tu.tz.21,lee.20}.
%By fitting the data using Eq. (\ref{scaling}), we obtain
%the central charge $c=-0.991$ and $c=-0.972$ for $\delta t=0.1+\epsilon$ and
%$\gamma=-(0.1+\epsilon)$, respectively. 
%The negative central charges cab be
%explained by the bc-GHOST CFT theory \cite{ch.yo.20}.
By fitting the data using Eq. (\ref{scaling}), we obtain $c=-1.3367$,
$s_1=-2.7697$ for $\delta t=0.1000001$ ($\epsilon=1\times 10^{-7}$),
while $c=-3.2026$, $s_1=3.1642$ for $\delta t=-0.10001$
($\epsilon=1\times 10^{-5}$). The values of $c$ and $s_1$ strongly
depend on the way one approaching the critical points from the
line-gap side \cite{ch.yo.20,tu.tz.21}, \textit{e.g.}, the magnitude of shift
$\epsilon$. For the critical point $\delta t=0.1$, the absolute value
$|c|$ seems to converge to a value around $1.34$ when
$\epsilon \rightarrow 0$. However, for the critical point
$\delta t=-0.1$, the value $|c|$ increases as $\epsilon \rightarrow 0$
and a violation of the universal relation in Eq. (\ref{scaling}) is
observed after $\epsilon <1\times 10^{-5}$.
When entering the region $-\gamma<\delta t<\gamma$, the scaling of
real part of EE crosses over from a concave ($c<0$) to convex ($c>0$)
function \cite{,ch.yo.20}; See Fig. \ref{fig2}(c) for examples, in
which $c=0.9767$, $s_1=0.8164$ and $c=0.9748$, $s_1=0.6105$ are
obtained for $\delta t=-0.05$ and $\delta t=0.05$, respectively. It is
interesting that the whole region of $-\gamma<\delta t<\gamma$ is
critical with long-range correlations \cite{he.re.19}. The point
$\delta t=0$ is a topological critical point, at which the EE is real
and the fitting central charge is $c=1.0007$ with $s_1=0.7252$.
%It also with the apperian of of OBC.

The edge entanglement entropy defined in Eq.(\ref{eq:Sedge}) is a
useful tool to detect the topological properties of both
noninteracting and interacting many-body systems
\cite{wa.xu.15,kim.14,fr.ma.20,le.le.20}. It measures the entanglement
between two edges of a finite chain and converges to edge degeneracy
in the thermodynamic limit \cite{wa.xu.15,ye.mu.16}.  In
Fig. \ref{fig3}, the second-order edge entanglement entropy
$S_{2 ,edge}$ of noninteracting SSH model is plotted as a function of
$\delta t$ for a fixed value of $\gamma=0.1$. We use the same scheme
as Ref. \cite{wa.xu.15} to cut the system: When $\delta t>0$, the
subsystem $A$ are chosen as $[1;L/2]$ for both PBC and OBC chains;
When $\delta t<0$, subsystem $A$ are chosen as $[1;L/2+1]$ for OBC
chain, but as $[2;L/2+1]$ for PBC chain. By this way, all the cuts are
at the weak bonds \cite{wa.xu.15}.

\begin{figure*}[htbp]
  \centering
  \includegraphics[width=0.96\linewidth]{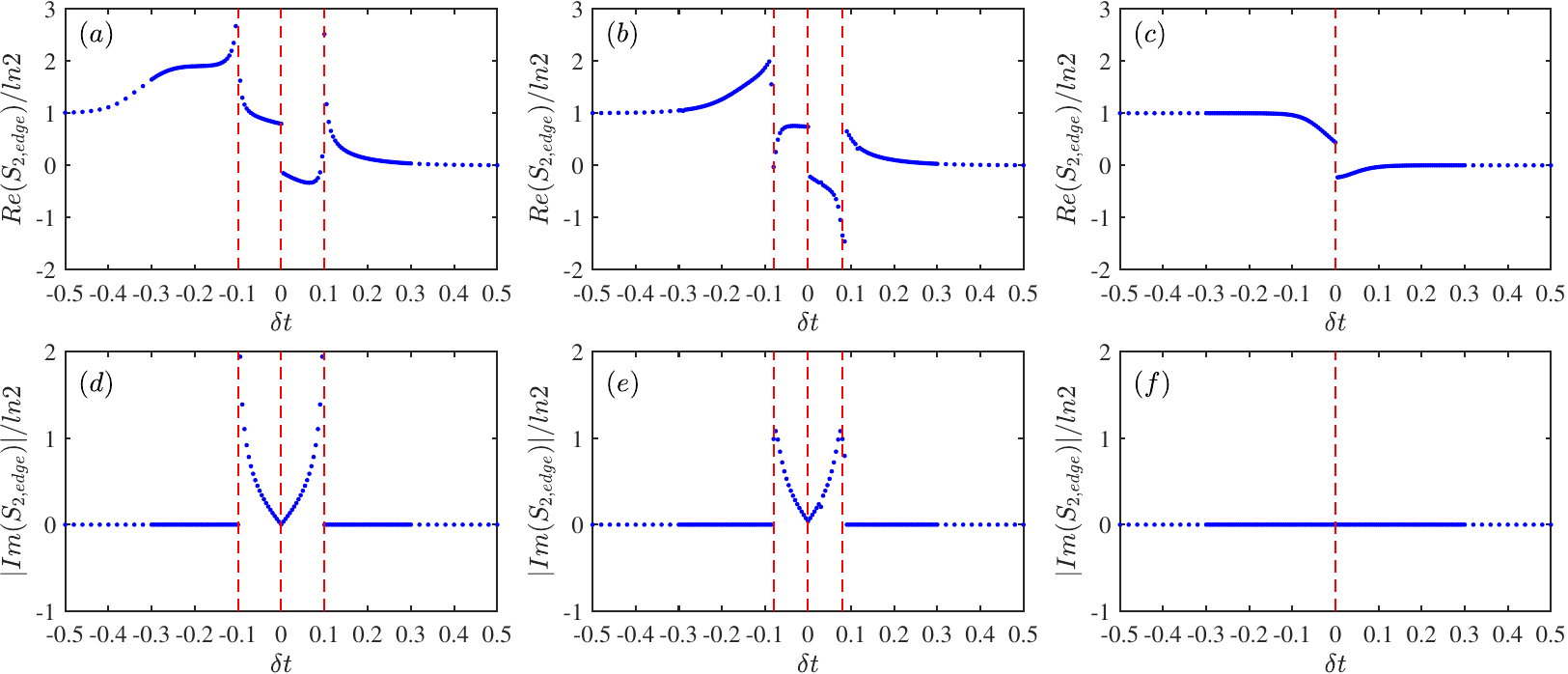}
  \caption{The real and imaginary parts of second order edge
    entanglement entropy $S_{2 ,edge}$ as a function of the hopping
    dimerization $\delta t$ for various interaction $U$. For (a) and
    (d), $U=0.1$; for (b) and (e), $U=1$; for (c) and (f), $U=10$.
    The system is at half-filling with $L=12$ and
    $\gamma=0.1$. The cut scheme used in the calculations is the same
    as that in Fig. \ref{fig3}.}
\label{fig4}
\end{figure*}

\begin{figure}[t]
    \centering
    \includegraphics[width=1\linewidth]{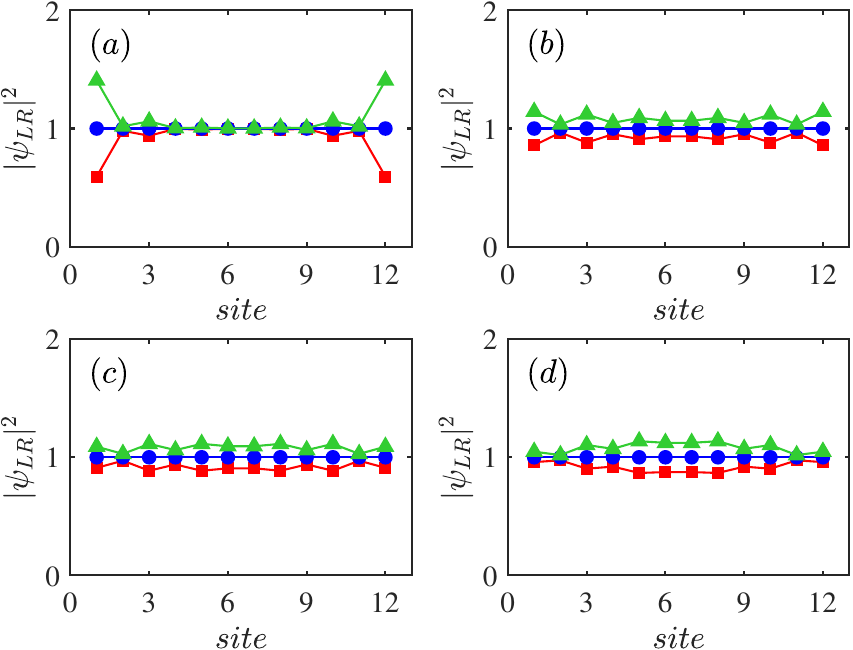}
    \caption{The distribution of particle density $|\psi_{LR}|^2$ of
      an open chain at finite interaction $U=1$. The blue dots
      correspond to the density at half-filling, while the green
      upward triangles (red square) are the density by adding
      (removing) one particle from the half-filling case. The values
      of parameters are $L=12$ and $\gamma=0.1$. From (a) to
      (d), $\delta t=-0.5$, $-0.08$, $0.08$, and $0.5$, respectively.}
    \label{fig5}
\end{figure}

%It is extend to the non-Hermitian case in Ref.[]
Again, four phase are recognized in Fig. \ref{fig3}: the line-gap
topological phase (phase $I$, $\delta t<-\gamma$), point-gap
topological phase (phase $II$, $-\gamma<\delta t<0$), point-gap
trivial phase (phase $III$, $0<\delta t<\gamma$), and line-gap trivial
phase (phase $IV$, $\delta t>\gamma$). In phases $I$ and $IV$, the
values of $S_{2 ,edge}$ are real and are quantized to $2\ln2$ and $0$,
respectively. Therefore, there are two degenerated edge modes in phase
$I$ and no edge modes in phase $IV$ \cite{le.le.20}, the same as the
topological and trivial phases of Hermitian SSH model
\cite{wa.xu.15}. In phases $II$ and $III$, the values of $S_{2 ,edge}$
are complex and are no longer quantized. The imaginary part of
$S_{2 ,edge}$ is rooted in the point-gap structure of energy spectrum,
an unique character of non-Hermitian systems.
%due to the point-gap structure, in which the
%conventional BBC is not well-defined. 
The value of $S_{2 ,edge}$ is
divergent as expected at the phase-transition points
$\delta t=\pm\gamma$ (gap-closing points), while shows a jump at
topological critical point $\delta t=0$.  
%This can be attributed to
%the appearance of edge states under OBC in the topological phase.  
In the topological phase, the edge states appear under OBC (also see
Fig. \ref{fig5}), which will
contribute a finite value to $S_{2 ,edge}$ and cause the jump.
As the lattice size $L$ increases, the peaks of $S_{2,edge}$ around
critical points $\delta t =\pm 0.1$ are narrowed, and the jump at
critical point $\delta t=0$ becomes smaller. Note that this jump
should decrease to zero in the thermodynamic limit
$L\rightarrow\infty$ for there is no difference between OBC and PBC
any longer.
An important feature of Fig. \ref{fig3} is that, even for a small
lattice size (\textit{e.g.}, $L=4$), the $S_{2 ,edge}$, which
includes information from both OBC and PBC chains, gives the correct
positions of three critical points. We therefore conclude that edge
entanglement entropy is a good indicator to discriminate different
phases of non-Hermitian systems.

%Note entire energy spectrum of our model is real in OBC \cite{le.le.20}

The edge entanglement entropy calculated by correlation matrix method
exactly matches that obtained by Lanczos ED method when $U=0$; see an
example of $L=12$ in Fig. \ref{fig3}. Note that there is a phase
random in the ground-state energy of Lanczos method. Due to the
time-reversal symmetry $T_{+}$ \cite{ka.sh.19}, the energy spectra of
Hamiltonian (\ref{model}) is composed of complex conjugate pairs $E_n$
and $E_n^*$. However, in the routines of Lanczos algorithm, the
eigenvalues pairs are not exactly complex conjugate each other, e.g.,
their real parts are tiny different. The ground state energy is chosen
according to the smallest real part, so that the sign of its imaginary
part is numerically "random" ($+$ or $-$). This results in the
"random" $\pm$ sign in the imaginary part of entanglement entropy. In
order to compare with the correlated matrix method, we therefore plot
the absolute value of imaginary part of $S_{2,edge}$ in
Fig. \ref{fig3}.

\section{Non-Hermitian Interacting SSH model}\label{sec4}

%for the formulation is based on many-boday basis, we can
%straithforward to generalize this into the interacting case....

We then move to study the BBC of non-Hermitian interacting SSHH model,
by examining the edge entanglement entropy computed using
the non-Hermitian ED method described in Sec. \ref{sec2}. The
numerical results of second-order edge entanglement entropy
$S_{2 ,edge}$ are presented in Fig. \ref{fig4} for various values of
interaction $U$.  When the $U$ is small, the overall behavior of
$S_{2,edge}$ is similar to the noninteracting case, except for the
quantized value in phase $I$ which decreases from $2\ln2$ to $\ln2$.
The Hubbard $U$ will raise the energy of doubly occupied edge state on
the right (or left) end of a open chain, and therefore reduce the edge
degeneracy 4 to 2, which results in the decrease of $S_{2 ,edge}$
\cite{wa.xu.15}. When $U$ is increasing as in Fig. \ref{fig4}(b), the
critical points on two sides move towards the center at $\delta t=0$,
\textit{i.e.}, the regions of non-Hermitian phases $II$ and $III$ are
narrowed and also the imaginary parts of $S_{2 ,edge}$ in them are
suppressed. For very large value of $U$ as in Fig. \ref{fig4}(c),
three critical points merge into the one at $\delta t=0$, and the
non-Hermitian phases $II$ and $III$ disappear in the phase diagram.
In this case, the system behaves like a Hermitian one with a real
value of $S_{2 ,edge}$, which suggests that the effects of
non-Hermiticity is killed by the interaction.

The particle density,
$|\psi_{LR}|^2=\langle \Psi_{L} |\sum_\sigma
n_{i,\sigma}|\Psi_{R}\rangle$, on each sites of a interacting chain
under OBC is shown in Fig. \ref{fig5}, for $U=1$ and four typical
values of $\delta t$.  At half-filling (blue dots), the distribution
of particle is homogeneous over the whole lattice, no matter what
values of $\delta t$ are. To detect the edge modes, we add or remove
one particle from the half-filling case \cite{le.le.20}. The additional particle
(hole) is localized near the edges when the system is in topologically
non-trivial phases ($\delta t<0$) as in Fig. \ref{fig5}(a) and (b).
The amplitude of edge mode in phase $II$ [Fig. \ref{fig5}(b)] is much 
smaller than that in phase $I$ due to the strong interference of non-Hermiticity,
and it will decrease to zero also when $\delta t$ approaching zero.
For the topologically trivial phases ($\delta t>0$), the
additional particle (hole) always spreads over the entire lattices, as
shown in Fig. \ref{fig5}(c) and (d), which means no edge modes in
these phases.

%When comparing with the noninteracting case, the ...
%We do not observed the anomalous boundary effect as shown in
%Ref. \cite{li.he.20} ,which due to the interplay of nonreciprocal
%hopping, superlattice potential,
%
%The behavior of the two
%        exceptional points of $ S_{2 ,edge}^{nH}$ is the same to the
%        first order entanglement spectrum, supporting the claim that
%        the interaction induces negative effect against
%        non-Hermiticity. 

\begin{figure}[!h]
	\centering
	\includegraphics[width=1\linewidth]{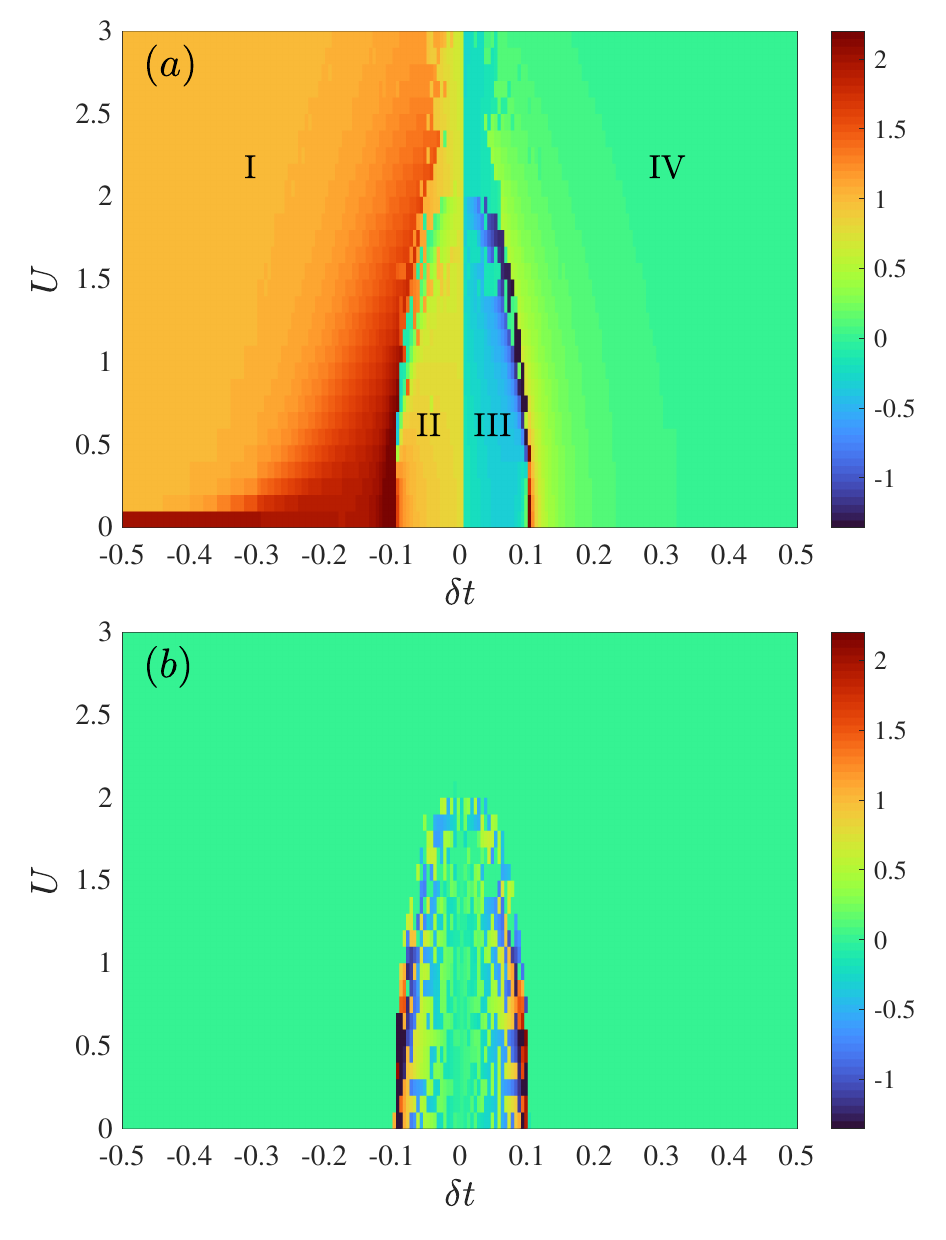}
	\caption{The phase diagram of non-Hermitian SSHH model in the
          plane of hopping dimeration $\delta t$ and interaction
          $U$. The system is at half-filling with $L=12$ and
          $\gamma=0.1$. Sub-figures (a) and (b) show the real and
          imaginary parts of second-order edge entanglement entropy
          $S_{2 ,edge}$ (in unit of $ln2$), respectively. }
	\label{fig6}
\end{figure}

To further reveal the details on the competition between the
interaction and the non-Hermiticity, we plot the comprehensive phase
diagram in Fig. \ref{fig6} in the plane of $\delta t$ and $U$,
according to the real and imaginary parts of second-order edge
entanglement entropy $S_{2 ,edge}$. The phase diagram can still be
roughly divided into four phases: the $S_{2 ,edge}$ is real for phases
$I$ and $IV$, while the imaginary part of $S_{2 ,edge}$ is nonzero for
phases $II$ and $III$, due to the different gap structure of energy
spectrum. As the increase of $U$, the point-gap region, characterized
by the imaginary part of $S_{2 ,edge}$ (phases $II$ and $III$ ), is
narrowing and eventually disappears around $U\approx 2.0$, which forms
a dome shape in the phase diagram. Note that the boundary of this dome
is blurred and irregular, which may due to 
%the finite size effect of calculations. 
the large quantum fluctuations here.
When $U$ is large, the quantized value of $S_{2,edge}$
is robust, \textit{i.e.}, $S_{2,edge}$ equals to $ln 2$ and $0$ in
phase $I$ and phase $IV$, respectively; see also
Fig. \ref{fig4}(c). There is only one topological transition point at
$\delta t=0$, implying that the behavior of system is similar to the
Hermitian one and the BBC recovers again.

%That is the BBC will disapperar by the interplay of interaction U.

The overall behaviors of phase boundaries with different
lattice size (\textit{e.g.}, $L=12$ and $L=8$) are similar, although
slightly differences appear in the intermediate $U$ region (close to
the critical interaction $U_c$) where the quantum fluctuations are
large. This implies that the finite-size effect is not severe in
determining the phase boundary according to $S_{2,edge}$. Note that it is
also difficult to compute $S_{2,edge}$ for a system with large $L$,
for one need to diagonalize a chain with OBC where the translational
invariance can not be utilized in Lanczos method.

The disappearance of non-Hermitian phases $II$ and $III$ in the phase
diagram can be understood from the viewpoint of Mott physics. For a
system at half-filling, Hubbard $U$ tends to localize the particles in
real space and pushes the system into the Mott-insulating phase, in
which the charge fluctuations (particle hopping) are greatly
suppressed. Therefore, the non-Hermitian phenomena associated with
hopping term can be suppressed also when the $U$ is increasing. 
Note that the physics would be significantly different if the
system is away from half-filling, due to the absence of Mott phase. 
%In general, the non-Hermitian effects, caused by the non-reciprocal
%hopping, are most striking at $U=0$. The presence of nonzero $U$ will
%lessen or destroy the non-Hermitian phenomena when the interaction
%term is dominant in the Hamiltonian. 
%This general behavior has been
%observed in other works also \cite{mu.le.20,yo.ha.21}.
The destruction of non-Hermitian effects by interaction has been
observed in some other works too \cite{mu.le.20,yo.ha.21}.  However,
in another context, the interaction can result in non-Hermiticity in
an original Hermitian system: the effective one-body quasiparticle
Hamiltonian is non-Hermitian when the lifetimes of different type of
quasiparticles are different
\cite{ko.fu.17,na.qi.20,cr.bu.21,mi.me.21}. The origin of
non-Hermiticity in quasiparticle context is different from that in
this paper (non-reciprocal hopping). It will be interesting to study
the quasiparticle behavior of non-Hermitian SSHH model in
Eq. (\ref{model}), which may exhibit novel non-Hermitian effects. We
leave this topic for future study.

%In the other context, the interaction results in an effective one-body
%Hamiltonian for quasiparticle
%\cite{ko.fu.17,na.qi.20,cr.bu.21,mi.me.21}, although the original
%Hamitonian is Hermitian. The

\section{Summary}\label{sec5}

In summary, we generalize the concept of edge entanglement entropy to
characterize different phases in non-Hermitian many-body systems. The
effectiveness of edge entanglement entropy to detect the topolgoical
properties of non-Hermitian systems is firstly examined on the free
SSH model. We demonstrated that the edge entanglement entropy,
including both OBC and PBC informations, gives the correct positions
of all critical points, and therefore is a good indicator for the
breakdown of BBC of non-Hermitian systems. For the interacting
non-Hermitian systems, we use the non-Hermitian Lanczos method to
compute the entanglement entropy. A comprehensive phase diagram was
obtained for the non-Hermitian SSHH model, according to both real and
imaginary parts of the second-order edge entanglement entropy. The
interplay between non-Hermitian and Hubbard interaction is analyzed,
and four different phases are identified in the phase diagram. We
showed that the breakdown of BBC is diminished and weakened by the
Hubbard interaction, which eventually disappears when interaction is
large.

%===============================================================================

\begin{acknowledgments}
  We are grateful for helpful discussions with Tian-Sheng Zeng, Hang
  Zhou, and Ching Hua Lee. This work is supported by the National
  Natural Science Foundation of China (Grant No. 11974293,
  No. 11874187, and No. 12174168) and the Fundamental Research Funds
  for Central Universities (Grant No. 20720180015).

  Weitao Chen and Liangtao Peng contributed equally to this work.
\end{acknowledgments}

%===============================================================================
%\bibliographystyle{apsrev4-2}
%\bibliography{ref,/Users/xiancong/work/documents/bibliography/topological}

\begin{thebibliography}{113}%
\makeatletter
\providecommand \@ifxundefined [1]{%
 \@ifx{#1\undefined}
}%
\providecommand \@ifnum [1]{%
 \ifnum #1\expandafter \@firstoftwo
 \else \expandafter \@secondoftwo
 \fi
}%
\providecommand \@ifx [1]{%
 \ifx #1\expandafter \@firstoftwo
 \else \expandafter \@secondoftwo
 \fi
}%
\providecommand \natexlab [1]{#1}%
\providecommand \enquote  [1]{``#1''}%
\providecommand \bibnamefont  [1]{#1}%
\providecommand \bibfnamefont [1]{#1}%
\providecommand \citenamefont [1]{#1}%
\providecommand \href@noop [0]{\@secondoftwo}%
\providecommand \href [0]{\begingroup \@sanitize@url \@href}%
\providecommand \@href[1]{\@@startlink{#1}\@@href}%
\providecommand \@@href[1]{\endgroup#1\@@endlink}%
\providecommand \@sanitize@url [0]{\catcode `\\12\catcode `\$12\catcode
  `\&12\catcode `\#12\catcode `\^12\catcode `\_12\catcode `\%12\relax}%
\providecommand \@@startlink[1]{}%
\providecommand \@@endlink[0]{}%
\providecommand \url  [0]{\begingroup\@sanitize@url \@url }%
\providecommand \@url [1]{\endgroup\@href {#1}{\urlprefix }}%
\providecommand \urlprefix  [0]{URL }%
\providecommand \Eprint [0]{\href }%
\providecommand \doibase [0]{https://doi.org/}%
\providecommand \selectlanguage [0]{\@gobble}%
\providecommand \bibinfo  [0]{\@secondoftwo}%
\providecommand \bibfield  [0]{\@secondoftwo}%
\providecommand \translation [1]{[#1]}%
\providecommand \BibitemOpen [0]{}%
\providecommand \bibitemStop [0]{}%
\providecommand \bibitemNoStop [0]{.\EOS\space}%
\providecommand \EOS [0]{\spacefactor3000\relax}%
\providecommand \BibitemShut  [1]{\csname bibitem#1\endcsname}%
\let\auto@bib@innerbib\@empty
%</preamble>
\bibitem [{\citenamefont {Cao}\ and\ \citenamefont {Wiersig}(2015)}]{ca.wi.15}%
  \BibitemOpen
  \bibfield  {author} {\bibinfo {author} {\bibfnamefont {H.}~\bibnamefont
  {Cao}}\ and\ \bibinfo {author} {\bibfnamefont {J.}~\bibnamefont {Wiersig}},\
  }\href {https://doi.org/10.1103/RevModPhys.87.61} {\bibfield  {journal}
  {\bibinfo  {journal} {Rev. Mod. Phys.}\ }\textbf {\bibinfo {volume} {87}},\
  \bibinfo {pages} {61} (\bibinfo {year} {2015})}\BibitemShut {NoStop}%
\bibitem [{\citenamefont {Ding}\ \emph {et~al.}(2016)\citenamefont {Ding},
  \citenamefont {Ma}, \citenamefont {Xiao}, \citenamefont {Zhang},\ and\
  \citenamefont {Chan}}]{di.ma.16}%
  \BibitemOpen
  \bibfield  {author} {\bibinfo {author} {\bibfnamefont {K.}~\bibnamefont
  {Ding}}, \bibinfo {author} {\bibfnamefont {G.}~\bibnamefont {Ma}}, \bibinfo
  {author} {\bibfnamefont {M.}~\bibnamefont {Xiao}}, \bibinfo {author}
  {\bibfnamefont {Z.~Q.}\ \bibnamefont {Zhang}},\ and\ \bibinfo {author}
  {\bibfnamefont {C.~T.}\ \bibnamefont {Chan}},\ }\href
  {https://doi.org/10.1103/PhysRevX.6.021007} {\bibfield  {journal} {\bibinfo
  {journal} {Phys. Rev. X}\ }\textbf {\bibinfo {volume} {6}},\ \bibinfo {pages}
  {021007} (\bibinfo {year} {2016})}\BibitemShut {NoStop}%
\bibitem [{\citenamefont {Kawabata}\ \emph {et~al.}(2017)\citenamefont
  {Kawabata}, \citenamefont {Ashida},\ and\ \citenamefont {Ueda}}]{ka.as.17}%
  \BibitemOpen
  \bibfield  {author} {\bibinfo {author} {\bibfnamefont {K.}~\bibnamefont
  {Kawabata}}, \bibinfo {author} {\bibfnamefont {Y.}~\bibnamefont {Ashida}},\
  and\ \bibinfo {author} {\bibfnamefont {M.}~\bibnamefont {Ueda}},\ }\href
  {https://doi.org/10.1103/PhysRevLett.119.190401} {\bibfield  {journal}
  {\bibinfo  {journal} {Phys. Rev. Lett.}\ }\textbf {\bibinfo {volume} {119}},\
  \bibinfo {pages} {190401} (\bibinfo {year} {2017})}\BibitemShut {NoStop}%
\bibitem [{\citenamefont {Liertzer}\ \emph {et~al.}(2012)\citenamefont
  {Liertzer}, \citenamefont {Ge}, \citenamefont {Cerjan}, \citenamefont
  {Stone}, \citenamefont {T\"ureci},\ and\ \citenamefont {Rotter}}]{li.ge.12}%
  \BibitemOpen
  \bibfield  {author} {\bibinfo {author} {\bibfnamefont {M.}~\bibnamefont
  {Liertzer}}, \bibinfo {author} {\bibfnamefont {L.}~\bibnamefont {Ge}},
  \bibinfo {author} {\bibfnamefont {A.}~\bibnamefont {Cerjan}}, \bibinfo
  {author} {\bibfnamefont {A.~D.}\ \bibnamefont {Stone}}, \bibinfo {author}
  {\bibfnamefont {H.~E.}\ \bibnamefont {T\"ureci}},\ and\ \bibinfo {author}
  {\bibfnamefont {S.}~\bibnamefont {Rotter}},\ }\href
  {https://doi.org/10.1103/PhysRevLett.108.173901} {\bibfield  {journal}
  {\bibinfo  {journal} {Phys. Rev. Lett.}\ }\textbf {\bibinfo {volume} {108}},\
  \bibinfo {pages} {173901} (\bibinfo {year} {2012})}\BibitemShut {NoStop}%
\bibitem [{\citenamefont {Ozawa}\ \emph {et~al.}(2019)\citenamefont {Ozawa},
  \citenamefont {Price}, \citenamefont {Amo}, \citenamefont {Goldman},
  \citenamefont {Hafezi}, \citenamefont {Lu}, \citenamefont {Rechtsman},
  \citenamefont {Schuster}, \citenamefont {Simon}, \citenamefont {Zilberberg},\
  and\ \citenamefont {Carusotto}}]{oz.pr.19}%
  \BibitemOpen
  \bibfield  {author} {\bibinfo {author} {\bibfnamefont {T.}~\bibnamefont
  {Ozawa}}, \bibinfo {author} {\bibfnamefont {H.~M.}\ \bibnamefont {Price}},
  \bibinfo {author} {\bibfnamefont {A.}~\bibnamefont {Amo}}, \bibinfo {author}
  {\bibfnamefont {N.}~\bibnamefont {Goldman}}, \bibinfo {author} {\bibfnamefont
  {M.}~\bibnamefont {Hafezi}}, \bibinfo {author} {\bibfnamefont
  {L.}~\bibnamefont {Lu}}, \bibinfo {author} {\bibfnamefont {M.~C.}\
  \bibnamefont {Rechtsman}}, \bibinfo {author} {\bibfnamefont {D.}~\bibnamefont
  {Schuster}}, \bibinfo {author} {\bibfnamefont {J.}~\bibnamefont {Simon}},
  \bibinfo {author} {\bibfnamefont {O.}~\bibnamefont {Zilberberg}},\ and\
  \bibinfo {author} {\bibfnamefont {I.}~\bibnamefont {Carusotto}},\ }\href
  {https://doi.org/10.1103/RevModPhys.91.015006} {\bibfield  {journal}
  {\bibinfo  {journal} {Rev. Mod. Phys.}\ }\textbf {\bibinfo {volume} {91}},\
  \bibinfo {pages} {015006} (\bibinfo {year} {2019})}\BibitemShut {NoStop}%
\bibitem [{\citenamefont {Fleury}\ \emph {et~al.}(2015)\citenamefont {Fleury},
  \citenamefont {Sounas},\ and\ \citenamefont {AlÃ¹}}]{fl.so.15}%
  \BibitemOpen
  \bibfield  {author} {\bibinfo {author} {\bibfnamefont {R.}~\bibnamefont
  {Fleury}}, \bibinfo {author} {\bibfnamefont {D.}~\bibnamefont {Sounas}},\
  and\ \bibinfo {author} {\bibfnamefont {A.}~\bibnamefont {AlÃ¹}},\ }\href
  {https://doi.org/10.1038/ncomms6905} {\bibfield  {journal} {\bibinfo
  {journal} {Nature Communications}\ }\textbf {\bibinfo {volume} {6}},\
  \bibinfo {pages} {5905} (\bibinfo {year} {2015})}\BibitemShut {NoStop}%
\bibitem [{\citenamefont {Doppler}\ \emph {et~al.}(2016)\citenamefont
  {Doppler}, \citenamefont {Mailybaev}, \citenamefont {BÃ¶hm}, \citenamefont
  {Kuhl}, \citenamefont {Girschik}, \citenamefont {Libisch}, \citenamefont
  {Milburn}, \citenamefont {Rabl}, \citenamefont {Moiseyev},\ and\
  \citenamefont {Rotter}}]{do.ma.16}%
  \BibitemOpen
  \bibfield  {author} {\bibinfo {author} {\bibfnamefont {J.}~\bibnamefont
  {Doppler}}, \bibinfo {author} {\bibfnamefont {A.~A.}\ \bibnamefont
  {Mailybaev}}, \bibinfo {author} {\bibfnamefont {J.}~\bibnamefont {BÃ¶hm}},
  \bibinfo {author} {\bibfnamefont {U.}~\bibnamefont {Kuhl}}, \bibinfo {author}
  {\bibfnamefont {A.}~\bibnamefont {Girschik}}, \bibinfo {author}
  {\bibfnamefont {F.}~\bibnamefont {Libisch}}, \bibinfo {author} {\bibfnamefont
  {T.~J.}\ \bibnamefont {Milburn}}, \bibinfo {author} {\bibfnamefont
  {P.}~\bibnamefont {Rabl}}, \bibinfo {author} {\bibfnamefont {N.}~\bibnamefont
  {Moiseyev}},\ and\ \bibinfo {author} {\bibfnamefont {S.}~\bibnamefont
  {Rotter}},\ }\href {https://doi.org/10.1038/nature18605} {\bibfield
  {journal} {\bibinfo  {journal} {Nature}\ }\textbf {\bibinfo {volume} {537}},\
  \bibinfo {pages} {76} (\bibinfo {year} {2016})}\BibitemShut {NoStop}%
\bibitem [{\citenamefont {Malzard}\ \emph {et~al.}(2015)\citenamefont
  {Malzard}, \citenamefont {Poli},\ and\ \citenamefont {Schomerus}}]{ma.po.15}%
  \BibitemOpen
  \bibfield  {author} {\bibinfo {author} {\bibfnamefont {S.}~\bibnamefont
  {Malzard}}, \bibinfo {author} {\bibfnamefont {C.}~\bibnamefont {Poli}},\ and\
  \bibinfo {author} {\bibfnamefont {H.}~\bibnamefont {Schomerus}},\ }\href
  {https://doi.org/10.1103/PhysRevLett.115.200402} {\bibfield  {journal}
  {\bibinfo  {journal} {Phys. Rev. Lett.}\ }\textbf {\bibinfo {volume} {115}},\
  \bibinfo {pages} {200402} (\bibinfo {year} {2015})}\BibitemShut {NoStop}%
\bibitem [{\citenamefont {Chang}\ \emph {et~al.}(2014)\citenamefont {Chang},
  \citenamefont {Jiang}, \citenamefont {Hua}, \citenamefont {Yang},
  \citenamefont {Wen}, \citenamefont {Jiang}, \citenamefont {Li}, \citenamefont
  {Wang},\ and\ \citenamefont {Xiao}}]{ch.ji.14}%
  \BibitemOpen
  \bibfield  {author} {\bibinfo {author} {\bibfnamefont {L.}~\bibnamefont
  {Chang}}, \bibinfo {author} {\bibfnamefont {X.}~\bibnamefont {Jiang}},
  \bibinfo {author} {\bibfnamefont {S.}~\bibnamefont {Hua}}, \bibinfo {author}
  {\bibfnamefont {C.}~\bibnamefont {Yang}}, \bibinfo {author} {\bibfnamefont
  {J.}~\bibnamefont {Wen}}, \bibinfo {author} {\bibfnamefont {L.}~\bibnamefont
  {Jiang}}, \bibinfo {author} {\bibfnamefont {G.}~\bibnamefont {Li}}, \bibinfo
  {author} {\bibfnamefont {G.}~\bibnamefont {Wang}},\ and\ \bibinfo {author}
  {\bibfnamefont {M.}~\bibnamefont {Xiao}},\ }\href
  {https://doi.org/10.1038/nphoton.2014.133} {\bibfield  {journal} {\bibinfo
  {journal} {Nature Photonics}\ }\textbf {\bibinfo {volume} {8}},\ \bibinfo
  {pages} {524} (\bibinfo {year} {2014})}\BibitemShut {NoStop}%
\bibitem [{\citenamefont {Xu}\ \emph {et~al.}(2016)\citenamefont {Xu},
  \citenamefont {Mason}, \citenamefont {Jiang},\ and\ \citenamefont
  {Harris}}]{xu.ma.16}%
  \BibitemOpen
  \bibfield  {author} {\bibinfo {author} {\bibfnamefont {H.}~\bibnamefont
  {Xu}}, \bibinfo {author} {\bibfnamefont {D.}~\bibnamefont {Mason}}, \bibinfo
  {author} {\bibfnamefont {L.}~\bibnamefont {Jiang}},\ and\ \bibinfo {author}
  {\bibfnamefont {J.~G.~E.}\ \bibnamefont {Harris}},\ }\href
  {https://doi.org/10.1038/nature18604} {\bibfield  {journal} {\bibinfo
  {journal} {Nature}\ }\textbf {\bibinfo {volume} {537}},\ \bibinfo {pages}
  {80} (\bibinfo {year} {2016})}\BibitemShut {NoStop}%
\bibitem [{\citenamefont {Gao}\ \emph {et~al.}(2015)\citenamefont {Gao},
  \citenamefont {Estrecho}, \citenamefont {Bliokh}, \citenamefont {Liew},
  \citenamefont {Fraser}, \citenamefont {Brodbeck}, \citenamefont {Kamp},
  \citenamefont {Schneider}, \citenamefont {HÃ¶fling}, \citenamefont
  {Yamamoto}, \citenamefont {Nori}, \citenamefont {Kivshar}, \citenamefont
  {Truscott}, \citenamefont {Dall},\ and\ \citenamefont
  {Ostrovskaya}}]{ga.es.15}%
  \BibitemOpen
  \bibfield  {author} {\bibinfo {author} {\bibfnamefont {T.}~\bibnamefont
  {Gao}}, \bibinfo {author} {\bibfnamefont {E.}~\bibnamefont {Estrecho}},
  \bibinfo {author} {\bibfnamefont {K.~Y.}\ \bibnamefont {Bliokh}}, \bibinfo
  {author} {\bibfnamefont {T.~C.~H.}\ \bibnamefont {Liew}}, \bibinfo {author}
  {\bibfnamefont {M.~D.}\ \bibnamefont {Fraser}}, \bibinfo {author}
  {\bibfnamefont {S.}~\bibnamefont {Brodbeck}}, \bibinfo {author}
  {\bibfnamefont {M.}~\bibnamefont {Kamp}}, \bibinfo {author} {\bibfnamefont
  {C.}~\bibnamefont {Schneider}}, \bibinfo {author} {\bibfnamefont
  {S.}~\bibnamefont {HÃ¶fling}}, \bibinfo {author} {\bibfnamefont
  {Y.}~\bibnamefont {Yamamoto}}, \bibinfo {author} {\bibfnamefont
  {F.}~\bibnamefont {Nori}}, \bibinfo {author} {\bibfnamefont {Y.~S.}\
  \bibnamefont {Kivshar}}, \bibinfo {author} {\bibfnamefont {A.~G.}\
  \bibnamefont {Truscott}}, \bibinfo {author} {\bibfnamefont {R.~G.}\
  \bibnamefont {Dall}},\ and\ \bibinfo {author} {\bibfnamefont {E.~A.}\
  \bibnamefont {Ostrovskaya}},\ }\href {https://doi.org/10.1038/nature15522}
  {\bibfield  {journal} {\bibinfo  {journal} {Nature}\ }\textbf {\bibinfo
  {volume} {526}},\ \bibinfo {pages} {554} (\bibinfo {year}
  {2015})}\BibitemShut {NoStop}%
\bibitem [{\citenamefont {Makris}\ \emph {et~al.}(2008)\citenamefont {Makris},
  \citenamefont {El-Ganainy}, \citenamefont {Christodoulides},\ and\
  \citenamefont {Musslimani}}]{ma.el.08}%
  \BibitemOpen
  \bibfield  {author} {\bibinfo {author} {\bibfnamefont {K.~G.}\ \bibnamefont
  {Makris}}, \bibinfo {author} {\bibfnamefont {R.}~\bibnamefont {El-Ganainy}},
  \bibinfo {author} {\bibfnamefont {D.~N.}\ \bibnamefont {Christodoulides}},\
  and\ \bibinfo {author} {\bibfnamefont {Z.~H.}\ \bibnamefont {Musslimani}},\
  }\href {https://doi.org/10.1103/PhysRevLett.100.103904} {\bibfield  {journal}
  {\bibinfo  {journal} {Phys. Rev. Lett.}\ }\textbf {\bibinfo {volume} {100}},\
  \bibinfo {pages} {103904} (\bibinfo {year} {2008})}\BibitemShut {NoStop}%
\bibitem [{\citenamefont {Guo}\ \emph {et~al.}(2009)\citenamefont {Guo},
  \citenamefont {Salamo}, \citenamefont {Duchesne}, \citenamefont {Morandotti},
  \citenamefont {Volatier-Ravat}, \citenamefont {Aimez}, \citenamefont
  {Siviloglou},\ and\ \citenamefont {Christodoulides}}]{gu.sa.09}%
  \BibitemOpen
  \bibfield  {author} {\bibinfo {author} {\bibfnamefont {A.}~\bibnamefont
  {Guo}}, \bibinfo {author} {\bibfnamefont {G.~J.}\ \bibnamefont {Salamo}},
  \bibinfo {author} {\bibfnamefont {D.}~\bibnamefont {Duchesne}}, \bibinfo
  {author} {\bibfnamefont {R.}~\bibnamefont {Morandotti}}, \bibinfo {author}
  {\bibfnamefont {M.}~\bibnamefont {Volatier-Ravat}}, \bibinfo {author}
  {\bibfnamefont {V.}~\bibnamefont {Aimez}}, \bibinfo {author} {\bibfnamefont
  {G.~A.}\ \bibnamefont {Siviloglou}},\ and\ \bibinfo {author} {\bibfnamefont
  {D.~N.}\ \bibnamefont {Christodoulides}},\ }\href
  {https://doi.org/10.1103/PhysRevLett.103.093902} {\bibfield  {journal}
  {\bibinfo  {journal} {Phys. Rev. Lett.}\ }\textbf {\bibinfo {volume} {103}},\
  \bibinfo {pages} {093902} (\bibinfo {year} {2009})}\BibitemShut {NoStop}%
\bibitem [{\citenamefont {RÃ¼ter}\ \emph {et~al.}(2010)\citenamefont {RÃ¼ter},
  \citenamefont {Makris}, \citenamefont {El-Ganainy}, \citenamefont
  {Christodoulides}, \citenamefont {Segev},\ and\ \citenamefont
  {Kip}}]{ru.ma.10}%
  \BibitemOpen
  \bibfield  {author} {\bibinfo {author} {\bibfnamefont {C.~E.}\ \bibnamefont
  {RÃ¼ter}}, \bibinfo {author} {\bibfnamefont {K.~G.}\ \bibnamefont {Makris}},
  \bibinfo {author} {\bibfnamefont {R.}~\bibnamefont {El-Ganainy}}, \bibinfo
  {author} {\bibfnamefont {D.~N.}\ \bibnamefont {Christodoulides}}, \bibinfo
  {author} {\bibfnamefont {M.}~\bibnamefont {Segev}},\ and\ \bibinfo {author}
  {\bibfnamefont {D.}~\bibnamefont {Kip}},\ }\href
  {https://doi.org/10.1038/nphys1515} {\bibfield  {journal} {\bibinfo
  {journal} {Nature Physics}\ }\textbf {\bibinfo {volume} {6}},\ \bibinfo
  {pages} {192} (\bibinfo {year} {2010})}\BibitemShut {NoStop}%
\bibitem [{\citenamefont {Lin}\ \emph {et~al.}(2011)\citenamefont {Lin},
  \citenamefont {Ramezani}, \citenamefont {Eichelkraut}, \citenamefont
  {Kottos}, \citenamefont {Cao},\ and\ \citenamefont
  {Christodoulides}}]{li.Ra.11}%
  \BibitemOpen
  \bibfield  {author} {\bibinfo {author} {\bibfnamefont {Z.}~\bibnamefont
  {Lin}}, \bibinfo {author} {\bibfnamefont {H.}~\bibnamefont {Ramezani}},
  \bibinfo {author} {\bibfnamefont {T.}~\bibnamefont {Eichelkraut}}, \bibinfo
  {author} {\bibfnamefont {T.}~\bibnamefont {Kottos}}, \bibinfo {author}
  {\bibfnamefont {H.}~\bibnamefont {Cao}},\ and\ \bibinfo {author}
  {\bibfnamefont {D.~N.}\ \bibnamefont {Christodoulides}},\ }\href
  {https://doi.org/10.1103/PhysRevLett.106.213901} {\bibfield  {journal}
  {\bibinfo  {journal} {Phys. Rev. Lett.}\ }\textbf {\bibinfo {volume} {106}},\
  \bibinfo {pages} {213901} (\bibinfo {year} {2011})}\BibitemShut {NoStop}%
\bibitem [{\citenamefont {Lee}\ and\ \citenamefont {Chan}(2014)}]{le.ch.14}%
  \BibitemOpen
  \bibfield  {author} {\bibinfo {author} {\bibfnamefont {T.~E.}\ \bibnamefont
  {Lee}}\ and\ \bibinfo {author} {\bibfnamefont {C.-K.}\ \bibnamefont {Chan}},\
  }\href {https://doi.org/10.1103/PhysRevX.4.041001} {\bibfield  {journal}
  {\bibinfo  {journal} {Phys. Rev. X}\ }\textbf {\bibinfo {volume} {4}},\
  \bibinfo {pages} {041001} (\bibinfo {year} {2014})}\BibitemShut {NoStop}%
\bibitem [{\citenamefont {Ashida}\ \emph {et~al.}(2020)\citenamefont {Ashida},
  \citenamefont {Gong},\ and\ \citenamefont {Ueda}}]{as.go.20}%
  \BibitemOpen
  \bibfield  {author} {\bibinfo {author} {\bibfnamefont {Y.}~\bibnamefont
  {Ashida}}, \bibinfo {author} {\bibfnamefont {Z.}~\bibnamefont {Gong}},\ and\
  \bibinfo {author} {\bibfnamefont {M.}~\bibnamefont {Ueda}},\ }\href
  {https://doi.org/10.1080/00018732.2021.1876991} {\bibfield  {journal}
  {\bibinfo  {journal} {Advances in Physics}\ }\textbf {\bibinfo {volume}
  {69}},\ \bibinfo {pages} {249} (\bibinfo {year} {2020})}\BibitemShut
  {NoStop}%
\bibitem [{\citenamefont {Ma}\ and\ \citenamefont {Sheng}(2016)}]{ma.sh.16}%
  \BibitemOpen
  \bibfield  {author} {\bibinfo {author} {\bibfnamefont {G.}~\bibnamefont
  {Ma}}\ and\ \bibinfo {author} {\bibfnamefont {P.}~\bibnamefont {Sheng}},\
  }\href@noop {} {\bibfield  {journal} {\bibinfo  {journal} {Science advances}\
  }\textbf {\bibinfo {volume} {2}},\ \bibinfo {pages} {e1501595} (\bibinfo
  {year} {2016})}\BibitemShut {NoStop}%
\bibitem [{\citenamefont {Cummer}\ \emph {et~al.}(2016)\citenamefont {Cummer},
  \citenamefont {Christensen},\ and\ \citenamefont {AlÃ¹}}]{cu.ch.16}%
  \BibitemOpen
  \bibfield  {author} {\bibinfo {author} {\bibfnamefont {S.~A.}\ \bibnamefont
  {Cummer}}, \bibinfo {author} {\bibfnamefont {J.}~\bibnamefont
  {Christensen}},\ and\ \bibinfo {author} {\bibfnamefont {A.}~\bibnamefont
  {AlÃ¹}},\ }\href {https://doi.org/10.1038/natrevmats.2016.1} {\bibfield
  {journal} {\bibinfo  {journal} {Nature Reviews Materials}\ }\textbf {\bibinfo
  {volume} {1}},\ \bibinfo {pages} {16001} (\bibinfo {year}
  {2016})}\BibitemShut {NoStop}%
\bibitem [{\citenamefont {Zangeneh-Nejad}\ and\ \citenamefont
  {Fleury}(2019)}]{za.fl.19}%
  \BibitemOpen
  \bibfield  {author} {\bibinfo {author} {\bibfnamefont {F.}~\bibnamefont
  {Zangeneh-Nejad}}\ and\ \bibinfo {author} {\bibfnamefont {R.}~\bibnamefont
  {Fleury}},\ }\href
  {https://doi.org/https://doi.org/10.1016/j.revip.2019.100031} {\bibfield
  {journal} {\bibinfo  {journal} {Reviews in Physics}\ }\textbf {\bibinfo
  {volume} {4}},\ \bibinfo {pages} {100031} (\bibinfo {year}
  {2019})}\BibitemShut {NoStop}%
\bibitem [{\citenamefont {El-Ganainy}\ \emph {et~al.}(2018)\citenamefont
  {El-Ganainy}, \citenamefont {Makris}, \citenamefont {Khajavikhan},
  \citenamefont {Musslimani}, \citenamefont {Rotter},\ and\ \citenamefont
  {Christodoulides}}]{el.ma.18}%
  \BibitemOpen
  \bibfield  {author} {\bibinfo {author} {\bibfnamefont {R.}~\bibnamefont
  {El-Ganainy}}, \bibinfo {author} {\bibfnamefont {K.~G.}\ \bibnamefont
  {Makris}}, \bibinfo {author} {\bibfnamefont {M.}~\bibnamefont {Khajavikhan}},
  \bibinfo {author} {\bibfnamefont {Z.~H.}\ \bibnamefont {Musslimani}},
  \bibinfo {author} {\bibfnamefont {S.}~\bibnamefont {Rotter}},\ and\ \bibinfo
  {author} {\bibfnamefont {D.~N.}\ \bibnamefont {Christodoulides}},\ }\href
  {https://doi.org/10.1038/nphys4323} {\bibfield  {journal} {\bibinfo
  {journal} {Nature Physics}\ }\textbf {\bibinfo {volume} {14}},\ \bibinfo
  {pages} {11} (\bibinfo {year} {2018})}\BibitemShut {NoStop}%
\bibitem [{\citenamefont {Miri}\ and\ \citenamefont
  {Al{\`u}}(2019)}]{mi.al.19}%
  \BibitemOpen
  \bibfield  {author} {\bibinfo {author} {\bibfnamefont {M.-A.}\ \bibnamefont
  {Miri}}\ and\ \bibinfo {author} {\bibfnamefont {A.}~\bibnamefont {Al{\`u}}},\
  }\href {https://doi.org/10.1126/science.aar7709} {\bibfield  {journal}
  {\bibinfo  {journal} {Science}\ }\textbf {\bibinfo {volume} {363}},\ \bibinfo
  {pages} {eaar7709} (\bibinfo {year} {2019})}\BibitemShut {NoStop}%
\bibitem [{\citenamefont {Daley}(2014)}]{dale.14}%
  \BibitemOpen
  \bibfield  {author} {\bibinfo {author} {\bibfnamefont {A.~J.}\ \bibnamefont
  {Daley}},\ }\href {https://doi.org/10.1080/00018732.2014.933502} {\bibfield
  {journal} {\bibinfo  {journal} {Advances in Physics}\ }\textbf {\bibinfo
  {volume} {63}},\ \bibinfo {pages} {77} (\bibinfo {year} {2014})}\BibitemShut
  {NoStop}%
\bibitem [{\citenamefont {Kuhr}(2016)}]{kuhr.16}%
  \BibitemOpen
  \bibfield  {author} {\bibinfo {author} {\bibfnamefont {S.}~\bibnamefont
  {Kuhr}},\ }\href {https://doi.org/10.1093/nsr/nww023} {\bibfield  {journal}
  {\bibinfo  {journal} {National Science Review}\ }\textbf {\bibinfo {volume}
  {3}},\ \bibinfo {pages} {170} (\bibinfo {year} {2016})}\BibitemShut {NoStop}%
\bibitem [{\citenamefont {Li}\ \emph {et~al.}(2019)\citenamefont {Li},
  \citenamefont {Harter}, \citenamefont {Liu}, \citenamefont {de~Melo},
  \citenamefont {Joglekar},\ and\ \citenamefont {Luo}}]{li.ha.19}%
  \BibitemOpen
  \bibfield  {author} {\bibinfo {author} {\bibfnamefont {J.}~\bibnamefont
  {Li}}, \bibinfo {author} {\bibfnamefont {A.~K.}\ \bibnamefont {Harter}},
  \bibinfo {author} {\bibfnamefont {J.}~\bibnamefont {Liu}}, \bibinfo {author}
  {\bibfnamefont {L.}~\bibnamefont {de~Melo}}, \bibinfo {author} {\bibfnamefont
  {Y.~N.}\ \bibnamefont {Joglekar}},\ and\ \bibinfo {author} {\bibfnamefont
  {L.}~\bibnamefont {Luo}},\ }\href
  {https://doi.org/10.1038/s41467-019-08596-1} {\bibfield  {journal} {\bibinfo
  {journal} {Nature Communications}\ }\textbf {\bibinfo {volume} {10}},\
  \bibinfo {pages} {855} (\bibinfo {year} {2019})}\BibitemShut {NoStop}%
\bibitem [{\citenamefont {Lapp}\ \emph {et~al.}(2019)\citenamefont {Lapp},
  \citenamefont {Ang'ong'a}, \citenamefont {An},\ and\ \citenamefont
  {Gadway}}]{la.ja.19}%
  \BibitemOpen
  \bibfield  {author} {\bibinfo {author} {\bibfnamefont {S.}~\bibnamefont
  {Lapp}}, \bibinfo {author} {\bibfnamefont {J.}~\bibnamefont {Ang'ong'a}},
  \bibinfo {author} {\bibfnamefont {F.~A.}\ \bibnamefont {An}},\ and\ \bibinfo
  {author} {\bibfnamefont {B.}~\bibnamefont {Gadway}},\ }\href
  {https://doi.org/10.1088/1367-2630/ab1147} {\bibfield  {journal} {\bibinfo
  {journal} {New Journal of Physics}\ }\textbf {\bibinfo {volume} {21}},\
  \bibinfo {pages} {045006} (\bibinfo {year} {2019})}\BibitemShut {NoStop}%
\bibitem [{\citenamefont {Ren}\ \emph {et~al.}(2021)\citenamefont {Ren},
  \citenamefont {Liu}, \citenamefont {Zhao}, \citenamefont {He}, \citenamefont
  {Pak}, \citenamefont {Li},\ and\ \citenamefont {Jo}}]{re.li.21}%
  \BibitemOpen
  \bibfield  {author} {\bibinfo {author} {\bibfnamefont {Z.}~\bibnamefont
  {Ren}}, \bibinfo {author} {\bibfnamefont {D.}~\bibnamefont {Liu}}, \bibinfo
  {author} {\bibfnamefont {E.}~\bibnamefont {Zhao}}, \bibinfo {author}
  {\bibfnamefont {C.}~\bibnamefont {He}}, \bibinfo {author} {\bibfnamefont
  {K.~K.}\ \bibnamefont {Pak}}, \bibinfo {author} {\bibfnamefont
  {J.}~\bibnamefont {Li}},\ and\ \bibinfo {author} {\bibfnamefont {G.-B.}\
  \bibnamefont {Jo}},\ }\href@noop {} {\bibinfo {title} {Topological control of
  quantum states in non-hermitian spin-orbit-coupled fermions}} (\bibinfo
  {year} {2021}),\ \Eprint {https://arxiv.org/abs/2106.04874} {arXiv:2106.04874
  [cond-mat.quant-gas]} \BibitemShut {NoStop}%
\bibitem [{\citenamefont {Bender}(2007)}]{be.07}%
  \BibitemOpen
  \bibfield  {author} {\bibinfo {author} {\bibfnamefont {C.~M.}\ \bibnamefont
  {Bender}},\ }\href@noop {} {\bibfield  {journal} {\bibinfo  {journal}
  {Reports on Progress in Physics}\ }\textbf {\bibinfo {volume} {70}},\
  \bibinfo {pages} {947} (\bibinfo {year} {2007})}\BibitemShut {NoStop}%
\bibitem [{\citenamefont {Dorey}\ \emph {et~al.}(2007)\citenamefont {Dorey},
  \citenamefont {Dunning},\ and\ \citenamefont {Tateo}}]{do.du.07}%
  \BibitemOpen
  \bibfield  {author} {\bibinfo {author} {\bibfnamefont {P.}~\bibnamefont
  {Dorey}}, \bibinfo {author} {\bibfnamefont {C.}~\bibnamefont {Dunning}},\
  and\ \bibinfo {author} {\bibfnamefont {R.}~\bibnamefont {Tateo}},\
  }\href@noop {} {\bibfield  {journal} {\bibinfo  {journal} {Journal of Physics
  A: Mathematical and Theoretical}\ }\textbf {\bibinfo {volume} {40}},\
  \bibinfo {pages} {R205} (\bibinfo {year} {2007})}\BibitemShut {NoStop}%
\bibitem [{\citenamefont {Ãzdemir}\ \emph {et~al.}(2019)\citenamefont
  {Ãzdemir}, \citenamefont {Rotter}, \citenamefont {Nori},\ and\ \citenamefont
  {Yang}}]{oz.ro.19}%
  \BibitemOpen
  \bibfield  {author} {\bibinfo {author} {\bibfnamefont {Å.~K.}\ \bibnamefont
  {Ãzdemir}}, \bibinfo {author} {\bibfnamefont {S.}~\bibnamefont {Rotter}},
  \bibinfo {author} {\bibfnamefont {F.}~\bibnamefont {Nori}},\ and\ \bibinfo
  {author} {\bibfnamefont {L.}~\bibnamefont {Yang}},\ }\href
  {https://doi.org/10.1038/s41563-019-0304-9} {\bibfield  {journal} {\bibinfo
  {journal} {Nature Materials}\ }\textbf {\bibinfo {volume} {18}},\ \bibinfo
  {pages} {783} (\bibinfo {year} {2019})}\BibitemShut {NoStop}%
\bibitem [{\citenamefont {Deng}\ \emph {et~al.}(2010)\citenamefont {Deng},
  \citenamefont {Haug},\ and\ \citenamefont {Yamamoto}}]{de.ha.10}%
  \BibitemOpen
  \bibfield  {author} {\bibinfo {author} {\bibfnamefont {H.}~\bibnamefont
  {Deng}}, \bibinfo {author} {\bibfnamefont {H.}~\bibnamefont {Haug}},\ and\
  \bibinfo {author} {\bibfnamefont {Y.}~\bibnamefont {Yamamoto}},\ }\href
  {https://doi.org/10.1103/RevModPhys.82.1489} {\bibfield  {journal} {\bibinfo
  {journal} {Rev. Mod. Phys.}\ }\textbf {\bibinfo {volume} {82}},\ \bibinfo
  {pages} {1489} (\bibinfo {year} {2010})}\BibitemShut {NoStop}%
\bibitem [{\citenamefont {Berman}\ \emph {et~al.}(2006)\citenamefont {Berman},
  \citenamefont {Lin},\ and\ \citenamefont {Arimondo}}]{be.li.06}%
  \BibitemOpen
  \bibfield  {author} {\bibinfo {author} {\bibfnamefont {P.~R.}\ \bibnamefont
  {Berman}}, \bibinfo {author} {\bibfnamefont {C.~C.}\ \bibnamefont {Lin}},\
  and\ \bibinfo {author} {\bibfnamefont {E.}~\bibnamefont {Arimondo}},\
  }\href@noop {} {\emph {\bibinfo {title} {Advances in Atomic, Molecular, and
  Optical Physics}}}\ (\bibinfo  {publisher} {Elsevier},\ \bibinfo {year}
  {2006})\BibitemShut {NoStop}%
\bibitem [{\citenamefont {Ritsch}\ \emph {et~al.}(2013)\citenamefont {Ritsch},
  \citenamefont {Domokos}, \citenamefont {Brennecke},\ and\ \citenamefont
  {Esslinger}}]{ri.do.13}%
  \BibitemOpen
  \bibfield  {author} {\bibinfo {author} {\bibfnamefont {H.}~\bibnamefont
  {Ritsch}}, \bibinfo {author} {\bibfnamefont {P.}~\bibnamefont {Domokos}},
  \bibinfo {author} {\bibfnamefont {F.}~\bibnamefont {Brennecke}},\ and\
  \bibinfo {author} {\bibfnamefont {T.}~\bibnamefont {Esslinger}},\ }\href
  {https://doi.org/10.1103/RevModPhys.85.553} {\bibfield  {journal} {\bibinfo
  {journal} {Rev. Mod. Phys.}\ }\textbf {\bibinfo {volume} {85}},\ \bibinfo
  {pages} {553} (\bibinfo {year} {2013})}\BibitemShut {NoStop}%
\bibitem [{\citenamefont {Sieberer}\ \emph {et~al.}(2016)\citenamefont
  {Sieberer}, \citenamefont {Buchhold},\ and\ \citenamefont
  {Diehl}}]{si.bu.16}%
  \BibitemOpen
  \bibfield  {author} {\bibinfo {author} {\bibfnamefont {L.~M.}\ \bibnamefont
  {Sieberer}}, \bibinfo {author} {\bibfnamefont {M.}~\bibnamefont {Buchhold}},\
  and\ \bibinfo {author} {\bibfnamefont {S.}~\bibnamefont {Diehl}},\
  }\href@noop {} {\bibfield  {journal} {\bibinfo  {journal} {Reports on
  Progress in Physics}\ }\textbf {\bibinfo {volume} {79}},\ \bibinfo {pages}
  {096001} (\bibinfo {year} {2016})}\BibitemShut {NoStop}%
\bibitem [{\citenamefont {Weimer}\ \emph {et~al.}(2021)\citenamefont {Weimer},
  \citenamefont {Kshetrimayum},\ and\ \citenamefont {Or\'us}}]{we.ks.21}%
  \BibitemOpen
  \bibfield  {author} {\bibinfo {author} {\bibfnamefont {H.}~\bibnamefont
  {Weimer}}, \bibinfo {author} {\bibfnamefont {A.}~\bibnamefont
  {Kshetrimayum}},\ and\ \bibinfo {author} {\bibfnamefont {R.}~\bibnamefont
  {Or\'us}},\ }\href {https://doi.org/10.1103/RevModPhys.93.015008} {\bibfield
  {journal} {\bibinfo  {journal} {Rev. Mod. Phys.}\ }\textbf {\bibinfo {volume}
  {93}},\ \bibinfo {pages} {015008} (\bibinfo {year} {2021})}\BibitemShut
  {NoStop}%
\bibitem [{\citenamefont {Bergholtz}\ and\ \citenamefont
  {Budich}(2019)}]{be.bu.19}%
  \BibitemOpen
  \bibfield  {author} {\bibinfo {author} {\bibfnamefont {E.~J.}\ \bibnamefont
  {Bergholtz}}\ and\ \bibinfo {author} {\bibfnamefont {J.~C.}\ \bibnamefont
  {Budich}},\ }\href {https://doi.org/10.1103/PhysRevResearch.1.012003}
  {\bibfield  {journal} {\bibinfo  {journal} {Phys. Rev. Research}\ }\textbf
  {\bibinfo {volume} {1}},\ \bibinfo {pages} {012003} (\bibinfo {year}
  {2019})}\BibitemShut {NoStop}%
\bibitem [{\citenamefont {Cayao}\ and\ \citenamefont
  {Black-Schaffer}(2021)}]{ca.bl.21}%
  \BibitemOpen
  \bibfield  {author} {\bibinfo {author} {\bibfnamefont {J.}~\bibnamefont
  {Cayao}}\ and\ \bibinfo {author} {\bibfnamefont {A.~M.}\ \bibnamefont
  {Black-Schaffer}},\ }\href@noop {} {\bibinfo {title} {Exceptional
  odd-frequency pairing in non-hermitian superconducting systems}} (\bibinfo
  {year} {2021}),\ \Eprint {https://arxiv.org/abs/2107.04445} {arXiv:2107.04445
  [cond-mat.supr-con]} \BibitemShut {NoStop}%
\bibitem [{\citenamefont {San-Jose}\ \emph {et~al.}(2016)\citenamefont
  {San-Jose}, \citenamefont {Cayao}, \citenamefont {Prada},\ and\ \citenamefont
  {Aguado}}]{sa.ca.16}%
  \BibitemOpen
  \bibfield  {author} {\bibinfo {author} {\bibfnamefont {P.}~\bibnamefont
  {San-Jose}}, \bibinfo {author} {\bibfnamefont {J.}~\bibnamefont {Cayao}},
  \bibinfo {author} {\bibfnamefont {E.}~\bibnamefont {Prada}},\ and\ \bibinfo
  {author} {\bibfnamefont {R.}~\bibnamefont {Aguado}},\ }\href
  {https://doi.org/10.1038/srep21427} {\bibfield  {journal} {\bibinfo
  {journal} {Scientific Reports}\ }\textbf {\bibinfo {volume} {6}},\ \bibinfo
  {pages} {21427} (\bibinfo {year} {2016})}\BibitemShut {NoStop}%
\bibitem [{\citenamefont {Pikulin}\ and\ \citenamefont
  {Nazarov}(2013)}]{pi.na.13}%
  \BibitemOpen
  \bibfield  {author} {\bibinfo {author} {\bibfnamefont {D.~I.}\ \bibnamefont
  {Pikulin}}\ and\ \bibinfo {author} {\bibfnamefont {Y.~V.}\ \bibnamefont
  {Nazarov}},\ }\href {https://doi.org/10.1103/PhysRevB.87.235421} {\bibfield
  {journal} {\bibinfo  {journal} {Phys. Rev. B}\ }\textbf {\bibinfo {volume}
  {87}},\ \bibinfo {pages} {235421} (\bibinfo {year} {2013})}\BibitemShut
  {NoStop}%
\bibitem [{\citenamefont {Pikulin}\ and\ \citenamefont
  {Nazarov}(2012)}]{pi.na.12}%
  \BibitemOpen
  \bibfield  {author} {\bibinfo {author} {\bibfnamefont {D.~I.}\ \bibnamefont
  {Pikulin}}\ and\ \bibinfo {author} {\bibfnamefont {Y.~V.}\ \bibnamefont
  {Nazarov}},\ }\href {https://doi.org/10.1134/S0021364011210090} {\bibfield
  {journal} {\bibinfo  {journal} {JETP Letters}\ }\textbf {\bibinfo {volume}
  {94}},\ \bibinfo {pages} {693} (\bibinfo {year} {2012})}\BibitemShut
  {NoStop}%
\bibitem [{\citenamefont {Qi}\ and\ \citenamefont {Zhang}(2011)}]{qi.zh.11}%
  \BibitemOpen
  \bibfield  {author} {\bibinfo {author} {\bibfnamefont {X.-L.}\ \bibnamefont
  {Qi}}\ and\ \bibinfo {author} {\bibfnamefont {S.-C.}\ \bibnamefont {Zhang}},\
  }\href {https://doi.org/10.1103/RevModPhys.83.1057} {\bibfield  {journal}
  {\bibinfo  {journal} {Rev. Mod. Phys.}\ }\textbf {\bibinfo {volume} {83}},\
  \bibinfo {pages} {1057} (\bibinfo {year} {2011})}\BibitemShut {NoStop}%
\bibitem [{\citenamefont {Hasan}\ and\ \citenamefont {Kane}(2010)}]{ha.ka.10}%
  \BibitemOpen
  \bibfield  {author} {\bibinfo {author} {\bibfnamefont {M.~Z.}\ \bibnamefont
  {Hasan}}\ and\ \bibinfo {author} {\bibfnamefont {C.~L.}\ \bibnamefont
  {Kane}},\ }\href {https://doi.org/10.1103/RevModPhys.82.3045} {\bibfield
  {journal} {\bibinfo  {journal} {Rev. Mod. Phys.}\ }\textbf {\bibinfo {volume}
  {82}},\ \bibinfo {pages} {3045} (\bibinfo {year} {2010})}\BibitemShut
  {NoStop}%
\bibitem [{\citenamefont {Ryu}\ and\ \citenamefont
  {Hatsugai}(2006)}]{ry.ha.06}%
  \BibitemOpen
  \bibfield  {author} {\bibinfo {author} {\bibfnamefont {S.}~\bibnamefont
  {Ryu}}\ and\ \bibinfo {author} {\bibfnamefont {Y.}~\bibnamefont {Hatsugai}},\
  }\href {https://doi.org/10.1103/PhysRevB.73.245115} {\bibfield  {journal}
  {\bibinfo  {journal} {Phys. Rev. B}\ }\textbf {\bibinfo {volume} {73}},\
  \bibinfo {pages} {245115} (\bibinfo {year} {2006})}\BibitemShut {NoStop}%
\bibitem [{\citenamefont {Grusdt}\ \emph {et~al.}(2013)\citenamefont {Grusdt},
  \citenamefont {H\"oning},\ and\ \citenamefont {Fleischhauer}}]{gr.ho.13}%
  \BibitemOpen
  \bibfield  {author} {\bibinfo {author} {\bibfnamefont {F.}~\bibnamefont
  {Grusdt}}, \bibinfo {author} {\bibfnamefont {M.}~\bibnamefont {H\"oning}},\
  and\ \bibinfo {author} {\bibfnamefont {M.}~\bibnamefont {Fleischhauer}},\
  }\href {https://doi.org/10.1103/PhysRevLett.110.260405} {\bibfield  {journal}
  {\bibinfo  {journal} {Phys. Rev. Lett.}\ }\textbf {\bibinfo {volume} {110}},\
  \bibinfo {pages} {260405} (\bibinfo {year} {2013})}\BibitemShut {NoStop}%
\bibitem [{\citenamefont {Rhim}\ \emph {et~al.}(2017)\citenamefont {Rhim},
  \citenamefont {Behrends},\ and\ \citenamefont {Bardarson}}]{rh.be.17}%
  \BibitemOpen
  \bibfield  {author} {\bibinfo {author} {\bibfnamefont {J.-W.}\ \bibnamefont
  {Rhim}}, \bibinfo {author} {\bibfnamefont {J.}~\bibnamefont {Behrends}},\
  and\ \bibinfo {author} {\bibfnamefont {J.~H.}\ \bibnamefont {Bardarson}},\
  }\href {https://doi.org/10.1103/PhysRevB.95.035421} {\bibfield  {journal}
  {\bibinfo  {journal} {Phys. Rev. B}\ }\textbf {\bibinfo {volume} {95}},\
  \bibinfo {pages} {035421} (\bibinfo {year} {2017})}\BibitemShut {NoStop}%
\bibitem [{\citenamefont {Lee}(2016)}]{lee.16}%
  \BibitemOpen
  \bibfield  {author} {\bibinfo {author} {\bibfnamefont {T.~E.}\ \bibnamefont
  {Lee}},\ }\href {https://doi.org/10.1103/PhysRevLett.116.133903} {\bibfield
  {journal} {\bibinfo  {journal} {Phys. Rev. Lett.}\ }\textbf {\bibinfo
  {volume} {116}},\ \bibinfo {pages} {133903} (\bibinfo {year}
  {2016})}\BibitemShut {NoStop}%
\bibitem [{\citenamefont {Xiong}(2018)}]{xion.18}%
  \BibitemOpen
  \bibfield  {author} {\bibinfo {author} {\bibfnamefont {Y.}~\bibnamefont
  {Xiong}},\ }\href@noop {} {\bibfield  {journal} {\bibinfo  {journal} {Journal
  of Physics Communications}\ }\textbf {\bibinfo {volume} {2}},\ \bibinfo
  {pages} {035043} (\bibinfo {year} {2018})}\BibitemShut {NoStop}%
\bibitem [{\citenamefont {Yao}\ and\ \citenamefont {Wang}(2018)}]{ya.wa.18}%
  \BibitemOpen
  \bibfield  {author} {\bibinfo {author} {\bibfnamefont {S.}~\bibnamefont
  {Yao}}\ and\ \bibinfo {author} {\bibfnamefont {Z.}~\bibnamefont {Wang}},\
  }\href {https://doi.org/10.1103/PhysRevLett.121.086803} {\bibfield  {journal}
  {\bibinfo  {journal} {Phys. Rev. Lett.}\ }\textbf {\bibinfo {volume} {121}},\
  \bibinfo {pages} {086803} (\bibinfo {year} {2018})}\BibitemShut {NoStop}%
\bibitem [{\citenamefont {Kunst}\ \emph {et~al.}(2018)\citenamefont {Kunst},
  \citenamefont {Edvardsson}, \citenamefont {Budich},\ and\ \citenamefont
  {Bergholtz}}]{ku.ed.18}%
  \BibitemOpen
  \bibfield  {author} {\bibinfo {author} {\bibfnamefont {F.~K.}\ \bibnamefont
  {Kunst}}, \bibinfo {author} {\bibfnamefont {E.}~\bibnamefont {Edvardsson}},
  \bibinfo {author} {\bibfnamefont {J.~C.}\ \bibnamefont {Budich}},\ and\
  \bibinfo {author} {\bibfnamefont {E.~J.}\ \bibnamefont {Bergholtz}},\ }\href
  {https://doi.org/10.1103/PhysRevLett.121.026808} {\bibfield  {journal}
  {\bibinfo  {journal} {Phys. Rev. Lett.}\ }\textbf {\bibinfo {volume} {121}},\
  \bibinfo {pages} {026808} (\bibinfo {year} {2018})}\BibitemShut {NoStop}%
\bibitem [{\citenamefont {Bergholtz}\ \emph {et~al.}(2021)\citenamefont
  {Bergholtz}, \citenamefont {Budich},\ and\ \citenamefont {Kunst}}]{be.bu.21}%
  \BibitemOpen
  \bibfield  {author} {\bibinfo {author} {\bibfnamefont {E.~J.}\ \bibnamefont
  {Bergholtz}}, \bibinfo {author} {\bibfnamefont {J.~C.}\ \bibnamefont
  {Budich}},\ and\ \bibinfo {author} {\bibfnamefont {F.~K.}\ \bibnamefont
  {Kunst}},\ }\href {https://doi.org/10.1103/RevModPhys.93.015005} {\bibfield
  {journal} {\bibinfo  {journal} {Rev. Mod. Phys.}\ }\textbf {\bibinfo {volume}
  {93}},\ \bibinfo {pages} {015005} (\bibinfo {year} {2021})}\BibitemShut
  {NoStop}%
\bibitem [{\citenamefont {Okuma}\ \emph {et~al.}(2020)\citenamefont {Okuma},
  \citenamefont {Kawabata}, \citenamefont {Shiozaki},\ and\ \citenamefont
  {Sato}}]{ok.ka.20}%
  \BibitemOpen
  \bibfield  {author} {\bibinfo {author} {\bibfnamefont {N.}~\bibnamefont
  {Okuma}}, \bibinfo {author} {\bibfnamefont {K.}~\bibnamefont {Kawabata}},
  \bibinfo {author} {\bibfnamefont {K.}~\bibnamefont {Shiozaki}},\ and\
  \bibinfo {author} {\bibfnamefont {M.}~\bibnamefont {Sato}},\ }\href
  {https://doi.org/10.1103/PhysRevLett.124.086801} {\bibfield  {journal}
  {\bibinfo  {journal} {Phys. Rev. Lett.}\ }\textbf {\bibinfo {volume} {124}},\
  \bibinfo {pages} {086801} (\bibinfo {year} {2020})}\BibitemShut {NoStop}%
\bibitem [{\citenamefont {Zhang}\ \emph
  {et~al.}(2020{\natexlab{a}})\citenamefont {Zhang}, \citenamefont {Yang},\
  and\ \citenamefont {Fang}}]{zh.ya.20}%
  \BibitemOpen
  \bibfield  {author} {\bibinfo {author} {\bibfnamefont {K.}~\bibnamefont
  {Zhang}}, \bibinfo {author} {\bibfnamefont {Z.}~\bibnamefont {Yang}},\ and\
  \bibinfo {author} {\bibfnamefont {C.}~\bibnamefont {Fang}},\ }\href
  {https://doi.org/10.1103/PhysRevLett.125.126402} {\bibfield  {journal}
  {\bibinfo  {journal} {Phys. Rev. Lett.}\ }\textbf {\bibinfo {volume} {125}},\
  \bibinfo {pages} {126402} (\bibinfo {year} {2020}{\natexlab{a}})}\BibitemShut
  {NoStop}%
\bibitem [{\citenamefont {Borgnia}\ \emph {et~al.}(2020)\citenamefont
  {Borgnia}, \citenamefont {Kruchkov},\ and\ \citenamefont
  {Slager}}]{bo.kr.20}%
  \BibitemOpen
  \bibfield  {author} {\bibinfo {author} {\bibfnamefont {D.~S.}\ \bibnamefont
  {Borgnia}}, \bibinfo {author} {\bibfnamefont {A.~J.}\ \bibnamefont
  {Kruchkov}},\ and\ \bibinfo {author} {\bibfnamefont {R.-J.}\ \bibnamefont
  {Slager}},\ }\href {https://doi.org/10.1103/PhysRevLett.124.056802}
  {\bibfield  {journal} {\bibinfo  {journal} {Phys. Rev. Lett.}\ }\textbf
  {\bibinfo {volume} {124}},\ \bibinfo {pages} {056802} (\bibinfo {year}
  {2020})}\BibitemShut {NoStop}%
\bibitem [{\citenamefont {Jin}\ and\ \citenamefont {Song}(2019)}]{ji.so.19}%
  \BibitemOpen
  \bibfield  {author} {\bibinfo {author} {\bibfnamefont {L.}~\bibnamefont
  {Jin}}\ and\ \bibinfo {author} {\bibfnamefont {Z.}~\bibnamefont {Song}},\
  }\href {https://doi.org/10.1103/PhysRevB.99.081103} {\bibfield  {journal}
  {\bibinfo  {journal} {Phys. Rev. B}\ }\textbf {\bibinfo {volume} {99}},\
  \bibinfo {pages} {081103} (\bibinfo {year} {2019})}\BibitemShut {NoStop}%
\bibitem [{\citenamefont {Zirnstein}\ \emph {et~al.}(2021)\citenamefont
  {Zirnstein}, \citenamefont {Refael},\ and\ \citenamefont
  {Rosenow}}]{zi.re.21}%
  \BibitemOpen
  \bibfield  {author} {\bibinfo {author} {\bibfnamefont {H.-G.}\ \bibnamefont
  {Zirnstein}}, \bibinfo {author} {\bibfnamefont {G.}~\bibnamefont {Refael}},\
  and\ \bibinfo {author} {\bibfnamefont {B.}~\bibnamefont {Rosenow}},\ }\href
  {https://doi.org/10.1103/PhysRevLett.126.216407} {\bibfield  {journal}
  {\bibinfo  {journal} {Phys. Rev. Lett.}\ }\textbf {\bibinfo {volume} {126}},\
  \bibinfo {pages} {216407} (\bibinfo {year} {2021})}\BibitemShut {NoStop}%
\bibitem [{\citenamefont {Yokomizo}\ and\ \citenamefont
  {Murakami}(2019)}]{yo.mu.19}%
  \BibitemOpen
  \bibfield  {author} {\bibinfo {author} {\bibfnamefont {K.}~\bibnamefont
  {Yokomizo}}\ and\ \bibinfo {author} {\bibfnamefont {S.}~\bibnamefont
  {Murakami}},\ }\href {https://doi.org/10.1103/PhysRevLett.123.066404}
  {\bibfield  {journal} {\bibinfo  {journal} {Phys. Rev. Lett.}\ }\textbf
  {\bibinfo {volume} {123}},\ \bibinfo {pages} {066404} (\bibinfo {year}
  {2019})}\BibitemShut {NoStop}%
\bibitem [{\citenamefont {Lee}\ and\ \citenamefont {Thomale}(2019)}]{le.th.19}%
  \BibitemOpen
  \bibfield  {author} {\bibinfo {author} {\bibfnamefont {C.~H.}\ \bibnamefont
  {Lee}}\ and\ \bibinfo {author} {\bibfnamefont {R.}~\bibnamefont {Thomale}},\
  }\href {https://doi.org/10.1103/PhysRevB.99.201103} {\bibfield  {journal}
  {\bibinfo  {journal} {Phys. Rev. B}\ }\textbf {\bibinfo {volume} {99}},\
  \bibinfo {pages} {201103} (\bibinfo {year} {2019})}\BibitemShut {NoStop}%
\bibitem [{\citenamefont {Imura}\ and\ \citenamefont
  {Takane}(2019)}]{im.ta.19}%
  \BibitemOpen
  \bibfield  {author} {\bibinfo {author} {\bibfnamefont {K.-I.}\ \bibnamefont
  {Imura}}\ and\ \bibinfo {author} {\bibfnamefont {Y.}~\bibnamefont {Takane}},\
  }\href {https://doi.org/10.1103/PhysRevB.100.165430} {\bibfield  {journal}
  {\bibinfo  {journal} {Phys. Rev. B}\ }\textbf {\bibinfo {volume} {100}},\
  \bibinfo {pages} {165430} (\bibinfo {year} {2019})}\BibitemShut {NoStop}%
\bibitem [{\citenamefont {Yang}\ \emph {et~al.}(2020)\citenamefont {Yang},
  \citenamefont {Zhang}, \citenamefont {Fang},\ and\ \citenamefont
  {Hu}}]{ya.zh.20}%
  \BibitemOpen
  \bibfield  {author} {\bibinfo {author} {\bibfnamefont {Z.}~\bibnamefont
  {Yang}}, \bibinfo {author} {\bibfnamefont {K.}~\bibnamefont {Zhang}},
  \bibinfo {author} {\bibfnamefont {C.}~\bibnamefont {Fang}},\ and\ \bibinfo
  {author} {\bibfnamefont {J.}~\bibnamefont {Hu}},\ }\href
  {https://doi.org/10.1103/PhysRevLett.125.226402} {\bibfield  {journal}
  {\bibinfo  {journal} {Phys. Rev. Lett.}\ }\textbf {\bibinfo {volume} {125}},\
  \bibinfo {pages} {226402} (\bibinfo {year} {2020})}\BibitemShut {NoStop}%
\bibitem [{\citenamefont {Herviou}\ \emph {et~al.}(2019)\citenamefont
  {Herviou}, \citenamefont {Bardarson},\ and\ \citenamefont
  {Regnault}}]{he.ba.19}%
  \BibitemOpen
  \bibfield  {author} {\bibinfo {author} {\bibfnamefont {L.}~\bibnamefont
  {Herviou}}, \bibinfo {author} {\bibfnamefont {J.~H.}\ \bibnamefont
  {Bardarson}},\ and\ \bibinfo {author} {\bibfnamefont {N.}~\bibnamefont
  {Regnault}},\ }\href {https://doi.org/10.1103/PhysRevA.99.052118} {\bibfield
  {journal} {\bibinfo  {journal} {Phys. Rev. A}\ }\textbf {\bibinfo {volume}
  {99}},\ \bibinfo {pages} {052118} (\bibinfo {year} {2019})}\BibitemShut
  {NoStop}%
\bibitem [{\citenamefont {Mu}\ \emph {et~al.}(2020)\citenamefont {Mu},
  \citenamefont {Lee}, \citenamefont {Li},\ and\ \citenamefont
  {Gong}}]{mu.le.20}%
  \BibitemOpen
  \bibfield  {author} {\bibinfo {author} {\bibfnamefont {S.}~\bibnamefont
  {Mu}}, \bibinfo {author} {\bibfnamefont {C.~H.}\ \bibnamefont {Lee}},
  \bibinfo {author} {\bibfnamefont {L.}~\bibnamefont {Li}},\ and\ \bibinfo
  {author} {\bibfnamefont {J.}~\bibnamefont {Gong}},\ }\href
  {https://doi.org/10.1103/PhysRevB.102.081115} {\bibfield  {journal} {\bibinfo
   {journal} {Phys. Rev. B}\ }\textbf {\bibinfo {volume} {102}},\ \bibinfo
  {pages} {081115} (\bibinfo {year} {2020})}\BibitemShut {NoStop}%
\bibitem [{\citenamefont {Lee}\ \emph {et~al.}(2020)\citenamefont {Lee},
  \citenamefont {Lee},\ and\ \citenamefont {Yang}}]{le.le.20}%
  \BibitemOpen
  \bibfield  {author} {\bibinfo {author} {\bibfnamefont {E.}~\bibnamefont
  {Lee}}, \bibinfo {author} {\bibfnamefont {H.}~\bibnamefont {Lee}},\ and\
  \bibinfo {author} {\bibfnamefont {B.-J.}\ \bibnamefont {Yang}},\ }\href
  {https://doi.org/10.1103/PhysRevB.101.121109} {\bibfield  {journal} {\bibinfo
   {journal} {Phys. Rev. B}\ }\textbf {\bibinfo {volume} {101}},\ \bibinfo
  {pages} {121109} (\bibinfo {year} {2020})}\BibitemShut {NoStop}%
\bibitem [{\citenamefont {Hamazaki}\ \emph {et~al.}(2019)\citenamefont
  {Hamazaki}, \citenamefont {Kawabata},\ and\ \citenamefont {Ueda}}]{ha.ka.19}%
  \BibitemOpen
  \bibfield  {author} {\bibinfo {author} {\bibfnamefont {R.}~\bibnamefont
  {Hamazaki}}, \bibinfo {author} {\bibfnamefont {K.}~\bibnamefont {Kawabata}},\
  and\ \bibinfo {author} {\bibfnamefont {M.}~\bibnamefont {Ueda}},\ }\href
  {https://doi.org/10.1103/PhysRevLett.123.090603} {\bibfield  {journal}
  {\bibinfo  {journal} {Phys. Rev. Lett.}\ }\textbf {\bibinfo {volume} {123}},\
  \bibinfo {pages} {090603} (\bibinfo {year} {2019})}\BibitemShut {NoStop}%
\bibitem [{\citenamefont {Zhang}\ \emph
  {et~al.}(2020{\natexlab{b}})\citenamefont {Zhang}, \citenamefont {Chen},
  \citenamefont {Zhang}, \citenamefont {Lang}, \citenamefont {Li},\ and\
  \citenamefont {Zhu}}]{zh.ch.20}%
  \BibitemOpen
  \bibfield  {author} {\bibinfo {author} {\bibfnamefont {D.-W.}\ \bibnamefont
  {Zhang}}, \bibinfo {author} {\bibfnamefont {Y.-L.}\ \bibnamefont {Chen}},
  \bibinfo {author} {\bibfnamefont {G.-Q.}\ \bibnamefont {Zhang}}, \bibinfo
  {author} {\bibfnamefont {L.-J.}\ \bibnamefont {Lang}}, \bibinfo {author}
  {\bibfnamefont {Z.}~\bibnamefont {Li}},\ and\ \bibinfo {author}
  {\bibfnamefont {S.-L.}\ \bibnamefont {Zhu}},\ }\href
  {https://doi.org/10.1103/PhysRevB.101.235150} {\bibfield  {journal} {\bibinfo
   {journal} {Phys. Rev. B}\ }\textbf {\bibinfo {volume} {101}},\ \bibinfo
  {pages} {235150} (\bibinfo {year} {2020}{\natexlab{b}})}\BibitemShut
  {NoStop}%
\bibitem [{\citenamefont {Xu}\ and\ \citenamefont {Chen}(2020)}]{xu.ch.20}%
  \BibitemOpen
  \bibfield  {author} {\bibinfo {author} {\bibfnamefont {Z.}~\bibnamefont
  {Xu}}\ and\ \bibinfo {author} {\bibfnamefont {S.}~\bibnamefont {Chen}},\
  }\href {https://doi.org/10.1103/PhysRevB.102.035153} {\bibfield  {journal}
  {\bibinfo  {journal} {Phys. Rev. B}\ }\textbf {\bibinfo {volume} {102}},\
  \bibinfo {pages} {035153} (\bibinfo {year} {2020})}\BibitemShut {NoStop}%
\bibitem [{\citenamefont {Liu}\ \emph {et~al.}(2020)\citenamefont {Liu},
  \citenamefont {He}, \citenamefont {Yoshida}, \citenamefont {Xiang},\ and\
  \citenamefont {Nori}}]{li.he.20}%
  \BibitemOpen
  \bibfield  {author} {\bibinfo {author} {\bibfnamefont {T.}~\bibnamefont
  {Liu}}, \bibinfo {author} {\bibfnamefont {J.~J.}\ \bibnamefont {He}},
  \bibinfo {author} {\bibfnamefont {T.}~\bibnamefont {Yoshida}}, \bibinfo
  {author} {\bibfnamefont {Z.-L.}\ \bibnamefont {Xiang}},\ and\ \bibinfo
  {author} {\bibfnamefont {F.}~\bibnamefont {Nori}},\ }\href
  {https://doi.org/10.1103/PhysRevB.102.235151} {\bibfield  {journal} {\bibinfo
   {journal} {Phys. Rev. B}\ }\textbf {\bibinfo {volume} {102}},\ \bibinfo
  {pages} {235151} (\bibinfo {year} {2020})}\BibitemShut {NoStop}%
\bibitem [{\citenamefont {Okuma}\ and\ \citenamefont
  {Sato}(2021{\natexlab{a}})}]{ok.sa.21b}%
  \BibitemOpen
  \bibfield  {author} {\bibinfo {author} {\bibfnamefont {N.}~\bibnamefont
  {Okuma}}\ and\ \bibinfo {author} {\bibfnamefont {M.}~\bibnamefont {Sato}},\
  }\href {https://doi.org/10.1103/PhysRevLett.126.176601} {\bibfield  {journal}
  {\bibinfo  {journal} {Phys. Rev. Lett.}\ }\textbf {\bibinfo {volume} {126}},\
  \bibinfo {pages} {176601} (\bibinfo {year} {2021}{\natexlab{a}})}\BibitemShut
  {NoStop}%
\bibitem [{\citenamefont {Yoshida}(2021)}]{yosh.21}%
  \BibitemOpen
  \bibfield  {author} {\bibinfo {author} {\bibfnamefont {T.}~\bibnamefont
  {Yoshida}},\ }\href {https://doi.org/10.1103/PhysRevB.103.125145} {\bibfield
  {journal} {\bibinfo  {journal} {Phys. Rev. B}\ }\textbf {\bibinfo {volume}
  {103}},\ \bibinfo {pages} {125145} (\bibinfo {year} {2021})}\BibitemShut
  {NoStop}%
\bibitem [{\citenamefont {Xi}\ \emph {et~al.}(2021)\citenamefont {Xi},
  \citenamefont {Zhang}, \citenamefont {Gu},\ and\ \citenamefont
  {Chen}}]{xi.zh.21}%
  \BibitemOpen
  \bibfield  {author} {\bibinfo {author} {\bibfnamefont {W.}~\bibnamefont
  {Xi}}, \bibinfo {author} {\bibfnamefont {Z.-H.}\ \bibnamefont {Zhang}},
  \bibinfo {author} {\bibfnamefont {Z.-C.}\ \bibnamefont {Gu}},\ and\ \bibinfo
  {author} {\bibfnamefont {W.-Q.}\ \bibnamefont {Chen}},\ }\href
  {https://doi.org/https://doi.org/10.1016/j.scib.2021.04.027} {\bibfield
  {journal} {\bibinfo  {journal} {Science Bulletin}\ }\textbf {\bibinfo
  {volume} {66}},\ \bibinfo {pages} {1731} (\bibinfo {year}
  {2021})}\BibitemShut {NoStop}%
\bibitem [{\citenamefont {Yoshida}\ and\ \citenamefont
  {Hatsugai}(2021)}]{yo.ha.21}%
  \BibitemOpen
  \bibfield  {author} {\bibinfo {author} {\bibfnamefont {T.}~\bibnamefont
  {Yoshida}}\ and\ \bibinfo {author} {\bibfnamefont {Y.}~\bibnamefont
  {Hatsugai}},\ }\href {https://doi.org/10.1103/PhysRevB.104.075106} {\bibfield
   {journal} {\bibinfo  {journal} {Phys. Rev. B}\ }\textbf {\bibinfo {volume}
  {104}},\ \bibinfo {pages} {075106} (\bibinfo {year} {2021})}\BibitemShut
  {NoStop}%
\bibitem [{\citenamefont {Zhang}\ and\ \citenamefont {Song}(2021)}]{zh.so.21}%
  \BibitemOpen
  \bibfield  {author} {\bibinfo {author} {\bibfnamefont {X.~Z.}\ \bibnamefont
  {Zhang}}\ and\ \bibinfo {author} {\bibfnamefont {Z.}~\bibnamefont {Song}},\
  }\href {https://doi.org/10.1103/PhysRevB.103.235153} {\bibfield  {journal}
  {\bibinfo  {journal} {Phys. Rev. B}\ }\textbf {\bibinfo {volume} {103}},\
  \bibinfo {pages} {235153} (\bibinfo {year} {2021})}\BibitemShut {NoStop}%
\bibitem [{\citenamefont {Shen}\ and\ \citenamefont {Lee}(2021)}]{sh.le.21}%
  \BibitemOpen
  \bibfield  {author} {\bibinfo {author} {\bibfnamefont {R.}~\bibnamefont
  {Shen}}\ and\ \bibinfo {author} {\bibfnamefont {C.~H.}\ \bibnamefont {Lee}},\
  }\href@noop {} {\bibinfo {title} {Non-hermitian skin clusters from strong
  interactions}} (\bibinfo {year} {2021}),\ \Eprint
  {https://arxiv.org/abs/2107.03414} {arXiv:2107.03414 [cond-mat.str-el]}
  \BibitemShut {NoStop}%
\bibitem [{\citenamefont {Kitaev}\ and\ \citenamefont
  {Preskill}(2006)}]{ki.pr.06}%
  \BibitemOpen
  \bibfield  {author} {\bibinfo {author} {\bibfnamefont {A.}~\bibnamefont
  {Kitaev}}\ and\ \bibinfo {author} {\bibfnamefont {J.}~\bibnamefont
  {Preskill}},\ }\href {https://doi.org/10.1103/PhysRevLett.96.110404}
  {\bibfield  {journal} {\bibinfo  {journal} {Phys. Rev. Lett.}\ }\textbf
  {\bibinfo {volume} {96}},\ \bibinfo {pages} {110404} (\bibinfo {year}
  {2006})}\BibitemShut {NoStop}%
\bibitem [{\citenamefont {Levin}\ and\ \citenamefont {Wen}(2006)}]{le.we.06}%
  \BibitemOpen
  \bibfield  {author} {\bibinfo {author} {\bibfnamefont {M.}~\bibnamefont
  {Levin}}\ and\ \bibinfo {author} {\bibfnamefont {X.-G.}\ \bibnamefont
  {Wen}},\ }\href {https://doi.org/10.1103/PhysRevLett.96.110405} {\bibfield
  {journal} {\bibinfo  {journal} {Phys. Rev. Lett.}\ }\textbf {\bibinfo
  {volume} {96}},\ \bibinfo {pages} {110405} (\bibinfo {year}
  {2006})}\BibitemShut {NoStop}%
\bibitem [{\citenamefont {Laflorencie}(2016)}]{lafl.16}%
  \BibitemOpen
  \bibfield  {author} {\bibinfo {author} {\bibfnamefont {N.}~\bibnamefont
  {Laflorencie}},\ }\href
  {https://doi.org/https://doi.org/10.1016/j.physrep.2016.06.008} {\bibfield
  {journal} {\bibinfo  {journal} {Physics Reports}\ }\textbf {\bibinfo {volume}
  {646}},\ \bibinfo {pages} {1} (\bibinfo {year} {2016})}\BibitemShut {NoStop}%
\bibitem [{\citenamefont {Zhang}\ \emph {et~al.}(2011)\citenamefont {Zhang},
  \citenamefont {Grover},\ and\ \citenamefont {Vishwanath}}]{zh.gr.11}%
  \BibitemOpen
  \bibfield  {author} {\bibinfo {author} {\bibfnamefont {Y.}~\bibnamefont
  {Zhang}}, \bibinfo {author} {\bibfnamefont {T.}~\bibnamefont {Grover}},\ and\
  \bibinfo {author} {\bibfnamefont {A.}~\bibnamefont {Vishwanath}},\ }\href
  {https://doi.org/10.1103/PhysRevLett.107.067202} {\bibfield  {journal}
  {\bibinfo  {journal} {Phys. Rev. Lett.}\ }\textbf {\bibinfo {volume} {107}},\
  \bibinfo {pages} {067202} (\bibinfo {year} {2011})}\BibitemShut {NoStop}%
\bibitem [{\citenamefont {Zhang}\ \emph {et~al.}(2012)\citenamefont {Zhang},
  \citenamefont {Grover}, \citenamefont {Turner}, \citenamefont {Oshikawa},\
  and\ \citenamefont {Vishwanath}}]{zh.gr.12}%
  \BibitemOpen
  \bibfield  {author} {\bibinfo {author} {\bibfnamefont {Y.}~\bibnamefont
  {Zhang}}, \bibinfo {author} {\bibfnamefont {T.}~\bibnamefont {Grover}},
  \bibinfo {author} {\bibfnamefont {A.}~\bibnamefont {Turner}}, \bibinfo
  {author} {\bibfnamefont {M.}~\bibnamefont {Oshikawa}},\ and\ \bibinfo
  {author} {\bibfnamefont {A.}~\bibnamefont {Vishwanath}},\ }\href
  {https://doi.org/10.1103/PhysRevB.85.235151} {\bibfield  {journal} {\bibinfo
  {journal} {Phys. Rev. B}\ }\textbf {\bibinfo {volume} {85}},\ \bibinfo
  {pages} {235151} (\bibinfo {year} {2012})}\BibitemShut {NoStop}%
\bibitem [{\citenamefont {Yan}\ \emph {et~al.}(2011)\citenamefont {Yan},
  \citenamefont {Huse},\ and\ \citenamefont {White}}]{ya.hu.11}%
  \BibitemOpen
  \bibfield  {author} {\bibinfo {author} {\bibfnamefont {S.}~\bibnamefont
  {Yan}}, \bibinfo {author} {\bibfnamefont {D.~A.}\ \bibnamefont {Huse}},\ and\
  \bibinfo {author} {\bibfnamefont {S.~R.}\ \bibnamefont {White}},\ }\href
  {https://doi.org/10.1126/science.1201080} {\bibfield  {journal} {\bibinfo
  {journal} {Science}\ }\textbf {\bibinfo {volume} {332}},\ \bibinfo {pages}
  {1173} (\bibinfo {year} {2011})}\BibitemShut {NoStop}%
\bibitem [{\citenamefont {Jiang}\ \emph
  {et~al.}(2012{\natexlab{a}})\citenamefont {Jiang}, \citenamefont {Wang},\
  and\ \citenamefont {Balents}}]{ji.wa.12}%
  \BibitemOpen
  \bibfield  {author} {\bibinfo {author} {\bibfnamefont {H.-C.}\ \bibnamefont
  {Jiang}}, \bibinfo {author} {\bibfnamefont {Z.}~\bibnamefont {Wang}},\ and\
  \bibinfo {author} {\bibfnamefont {L.}~\bibnamefont {Balents}},\ }\href
  {https://doi.org/10.1038/nphys2465} {\bibfield  {journal} {\bibinfo
  {journal} {Nature Physics}\ }\textbf {\bibinfo {volume} {8}},\ \bibinfo
  {pages} {902} (\bibinfo {year} {2012}{\natexlab{a}})}\BibitemShut {NoStop}%
\bibitem [{\citenamefont {Jiang}\ \emph
  {et~al.}(2012{\natexlab{b}})\citenamefont {Jiang}, \citenamefont {Yao},\ and\
  \citenamefont {Balents}}]{ji.ya.12}%
  \BibitemOpen
  \bibfield  {author} {\bibinfo {author} {\bibfnamefont {H.-C.}\ \bibnamefont
  {Jiang}}, \bibinfo {author} {\bibfnamefont {H.}~\bibnamefont {Yao}},\ and\
  \bibinfo {author} {\bibfnamefont {L.}~\bibnamefont {Balents}},\ }\href
  {https://doi.org/10.1103/PhysRevB.86.024424} {\bibfield  {journal} {\bibinfo
  {journal} {Phys. Rev. B}\ }\textbf {\bibinfo {volume} {86}},\ \bibinfo
  {pages} {024424} (\bibinfo {year} {2012}{\natexlab{b}})}\BibitemShut
  {NoStop}%
\bibitem [{\citenamefont {Hawking}\ \emph {et~al.}(2001)\citenamefont
  {Hawking}, \citenamefont {Maldacena},\ and\ \citenamefont
  {Strominger}}]{ha.ma.01}%
  \BibitemOpen
  \bibfield  {author} {\bibinfo {author} {\bibfnamefont {S.}~\bibnamefont
  {Hawking}}, \bibinfo {author} {\bibfnamefont {J.}~\bibnamefont {Maldacena}},\
  and\ \bibinfo {author} {\bibfnamefont {A.}~\bibnamefont {Strominger}},\
  }\href@noop {} {\bibfield  {journal} {\bibinfo  {journal} {Journal of High
  Energy Physics}\ }\textbf {\bibinfo {volume} {2001}},\ \bibinfo {pages} {001}
  (\bibinfo {year} {2001})}\BibitemShut {NoStop}%
\bibitem [{\citenamefont {Hastings}(2007)}]{hast.07}%
  \BibitemOpen
  \bibfield  {author} {\bibinfo {author} {\bibfnamefont {M.~B.}\ \bibnamefont
  {Hastings}},\ }\href {https://doi.org/10.1088/1742-5468/2007/08/p08024}
  {\bibfield  {journal} {\bibinfo  {journal} {Journal of Statistical Mechanics:
  Theory and Experiment}\ }\textbf {\bibinfo {volume} {2007}},\ \bibinfo
  {pages} {P08024} (\bibinfo {year} {2007})}\BibitemShut {NoStop}%
\bibitem [{\citenamefont {Wolf}\ \emph {et~al.}(2008)\citenamefont {Wolf},
  \citenamefont {Verstraete}, \citenamefont {Hastings},\ and\ \citenamefont
  {Cirac}}]{wo.ve.08}%
  \BibitemOpen
  \bibfield  {author} {\bibinfo {author} {\bibfnamefont {M.~M.}\ \bibnamefont
  {Wolf}}, \bibinfo {author} {\bibfnamefont {F.}~\bibnamefont {Verstraete}},
  \bibinfo {author} {\bibfnamefont {M.~B.}\ \bibnamefont {Hastings}},\ and\
  \bibinfo {author} {\bibfnamefont {J.~I.}\ \bibnamefont {Cirac}},\ }\href
  {https://doi.org/10.1103/PhysRevLett.100.070502} {\bibfield  {journal}
  {\bibinfo  {journal} {Phys. Rev. Lett.}\ }\textbf {\bibinfo {volume} {100}},\
  \bibinfo {pages} {070502} (\bibinfo {year} {2008})}\BibitemShut {NoStop}%
\bibitem [{\citenamefont {Holzhey}\ \emph {et~al.}(1994)\citenamefont
  {Holzhey}, \citenamefont {Larsen},\ and\ \citenamefont {Wilczek}}]{ho.la.94}%
  \BibitemOpen
  \bibfield  {author} {\bibinfo {author} {\bibfnamefont {C.}~\bibnamefont
  {Holzhey}}, \bibinfo {author} {\bibfnamefont {F.}~\bibnamefont {Larsen}},\
  and\ \bibinfo {author} {\bibfnamefont {F.}~\bibnamefont {Wilczek}},\
  }\href@noop {} {\bibfield  {journal} {\bibinfo  {journal} {Nuclear Physics
  B}\ }\textbf {\bibinfo {volume} {424}},\ \bibinfo {pages} {443} (\bibinfo
  {year} {1994})}\BibitemShut {NoStop}%
\bibitem [{\citenamefont {Calabrese}\ and\ \citenamefont
  {Cardy}(2009)}]{ca.ca.09}%
  \BibitemOpen
  \bibfield  {author} {\bibinfo {author} {\bibfnamefont {P.}~\bibnamefont
  {Calabrese}}\ and\ \bibinfo {author} {\bibfnamefont {J.}~\bibnamefont
  {Cardy}},\ }\href@noop {} {\bibfield  {journal} {\bibinfo  {journal} {Journal
  of Physics A: Mathematical and Theoretical}\ }\textbf {\bibinfo {volume}
  {42}},\ \bibinfo {pages} {504005} (\bibinfo {year} {2009})}\BibitemShut
  {NoStop}%
\bibitem [{\citenamefont {Chang}\ \emph {et~al.}(2020)\citenamefont {Chang},
  \citenamefont {You}, \citenamefont {Wen},\ and\ \citenamefont
  {Ryu}}]{ch.yo.20}%
  \BibitemOpen
  \bibfield  {author} {\bibinfo {author} {\bibfnamefont {P.-Y.}\ \bibnamefont
  {Chang}}, \bibinfo {author} {\bibfnamefont {J.-S.}\ \bibnamefont {You}},
  \bibinfo {author} {\bibfnamefont {X.}~\bibnamefont {Wen}},\ and\ \bibinfo
  {author} {\bibfnamefont {S.}~\bibnamefont {Ryu}},\ }\href
  {https://doi.org/10.1103/PhysRevResearch.2.033069} {\bibfield  {journal}
  {\bibinfo  {journal} {Phys. Rev. Research}\ }\textbf {\bibinfo {volume}
  {2}},\ \bibinfo {pages} {033069} (\bibinfo {year} {2020})}\BibitemShut
  {NoStop}%
\bibitem [{\citenamefont {Couvreur}\ \emph {et~al.}(2017)\citenamefont
  {Couvreur}, \citenamefont {Jacobsen},\ and\ \citenamefont
  {Saleur}}]{co.ja.17}%
  \BibitemOpen
  \bibfield  {author} {\bibinfo {author} {\bibfnamefont {R.}~\bibnamefont
  {Couvreur}}, \bibinfo {author} {\bibfnamefont {J.~L.}\ \bibnamefont
  {Jacobsen}},\ and\ \bibinfo {author} {\bibfnamefont {H.}~\bibnamefont
  {Saleur}},\ }\href {https://doi.org/10.1103/PhysRevLett.119.040601}
  {\bibfield  {journal} {\bibinfo  {journal} {Phys. Rev. Lett.}\ }\textbf
  {\bibinfo {volume} {119}},\ \bibinfo {pages} {040601} (\bibinfo {year}
  {2017})}\BibitemShut {NoStop}%
\bibitem [{\citenamefont {{Herviou}}\ \emph {et~al.}(2019)\citenamefont
  {{Herviou}}, \citenamefont {{Regnault}},\ and\ \citenamefont
  {{Bardarson}}}]{he.re.19}%
  \BibitemOpen
  \bibfield  {author} {\bibinfo {author} {\bibfnamefont {L.}~\bibnamefont
  {{Herviou}}}, \bibinfo {author} {\bibfnamefont {N.}~\bibnamefont
  {{Regnault}}},\ and\ \bibinfo {author} {\bibfnamefont {J.~H.}\ \bibnamefont
  {{Bardarson}}},\ }\href {https://doi.org/10.21468/SciPostPhys.7.5.069}
  {\bibfield  {journal} {\bibinfo  {journal} {SciPost Physics}\ }\textbf
  {\bibinfo {volume} {7}},\ \bibinfo {eid} {069} (\bibinfo {year}
  {2019})}\BibitemShut {NoStop}%
\bibitem [{\citenamefont {Chen}\ \emph {et~al.}(2021)\citenamefont {Chen},
  \citenamefont {Chen},\ and\ \citenamefont {Ye}}]{ch.ch.21}%
  \BibitemOpen
  \bibfield  {author} {\bibinfo {author} {\bibfnamefont {L.-M.}\ \bibnamefont
  {Chen}}, \bibinfo {author} {\bibfnamefont {S.~A.}\ \bibnamefont {Chen}},\
  and\ \bibinfo {author} {\bibfnamefont {P.}~\bibnamefont {Ye}},\ }\href
  {https://doi.org/10.21468/SciPostPhys.11.1.003} {\bibfield  {journal}
  {\bibinfo  {journal} {SciPost Phys.}\ }\textbf {\bibinfo {volume} {11}},\
  \bibinfo {pages} {3} (\bibinfo {year} {2021})}\BibitemShut {NoStop}%
\bibitem [{\citenamefont {Okuma}\ and\ \citenamefont
  {Sato}(2021{\natexlab{b}})}]{ok.sa.21a}%
  \BibitemOpen
  \bibfield  {author} {\bibinfo {author} {\bibfnamefont {N.}~\bibnamefont
  {Okuma}}\ and\ \bibinfo {author} {\bibfnamefont {M.}~\bibnamefont {Sato}},\
  }\href {https://doi.org/10.1103/PhysRevB.103.085428} {\bibfield  {journal}
  {\bibinfo  {journal} {Phys. Rev. B}\ }\textbf {\bibinfo {volume} {103}},\
  \bibinfo {pages} {085428} (\bibinfo {year} {2021}{\natexlab{b}})}\BibitemShut
  {NoStop}%
\bibitem [{\citenamefont {B\'acsi}\ and\ \citenamefont
  {D\'ora}(2021)}]{ba.do.21}%
  \BibitemOpen
  \bibfield  {author} {\bibinfo {author} {\bibfnamefont {A.}~\bibnamefont
  {B\'acsi}}\ and\ \bibinfo {author} {\bibfnamefont {B.}~\bibnamefont
  {D\'ora}},\ }\href {https://doi.org/10.1103/PhysRevB.103.085137} {\bibfield
  {journal} {\bibinfo  {journal} {Phys. Rev. B}\ }\textbf {\bibinfo {volume}
  {103}},\ \bibinfo {pages} {085137} (\bibinfo {year} {2021})}\BibitemShut
  {NoStop}%
\bibitem [{\citenamefont {Modak}\ and\ \citenamefont
  {Mandal}(2021)}]{mo.ma.21}%
  \BibitemOpen
  \bibfield  {author} {\bibinfo {author} {\bibfnamefont {R.}~\bibnamefont
  {Modak}}\ and\ \bibinfo {author} {\bibfnamefont {B.~P.}\ \bibnamefont
  {Mandal}},\ }\href {https://doi.org/10.1103/PhysRevA.103.062416} {\bibfield
  {journal} {\bibinfo  {journal} {Phys. Rev. A}\ }\textbf {\bibinfo {volume}
  {103}},\ \bibinfo {pages} {062416} (\bibinfo {year} {2021})}\BibitemShut
  {NoStop}%
\bibitem [{\citenamefont {Guo}\ \emph {et~al.}(2021)\citenamefont {Guo},
  \citenamefont {Yu}, \citenamefont {Huang}, \citenamefont {Yang},
  \citenamefont {Chi}, \citenamefont {Liao},\ and\ \citenamefont
  {Xiang}}]{gu.yu.21}%
  \BibitemOpen
  \bibfield  {author} {\bibinfo {author} {\bibfnamefont {Y.-B.}\ \bibnamefont
  {Guo}}, \bibinfo {author} {\bibfnamefont {Y.-C.}\ \bibnamefont {Yu}},
  \bibinfo {author} {\bibfnamefont {R.-Z.}\ \bibnamefont {Huang}}, \bibinfo
  {author} {\bibfnamefont {L.-P.}\ \bibnamefont {Yang}}, \bibinfo {author}
  {\bibfnamefont {R.-Z.}\ \bibnamefont {Chi}}, \bibinfo {author} {\bibfnamefont
  {H.-J.}\ \bibnamefont {Liao}},\ and\ \bibinfo {author} {\bibfnamefont
  {T.}~\bibnamefont {Xiang}},\ }\href
  {https://doi.org/10.1088/1361-648x/ac216e} {\bibfield  {journal} {\bibinfo
  {journal} {J. Phys.: Condens. Matter}\ }\textbf {\bibinfo {volume} {33}},\
  \bibinfo {pages} {475502} (\bibinfo {year} {2021})}\BibitemShut {NoStop}%
\bibitem [{\citenamefont {Tu}\ \emph {et~al.}(2021)\citenamefont {Tu},
  \citenamefont {Tzeng},\ and\ \citenamefont {Chang}}]{tu.tz.21}%
  \BibitemOpen
  \bibfield  {author} {\bibinfo {author} {\bibfnamefont {Y.-T.}\ \bibnamefont
  {Tu}}, \bibinfo {author} {\bibfnamefont {Y.-C.}\ \bibnamefont {Tzeng}},\ and\
  \bibinfo {author} {\bibfnamefont {P.-Y.}\ \bibnamefont {Chang}},\ }\href@noop
  {} {\bibinfo {title} {R\'enyi entropies and negative central charges in
  non-hermitian quantum systems}} (\bibinfo {year} {2021}),\ \Eprint
  {https://arxiv.org/abs/2107.13006} {arXiv:2107.13006 [cond-mat.str-el]}
  \BibitemShut {NoStop}%
\bibitem [{\citenamefont {Lee}(2020)}]{lee.20}%
  \BibitemOpen
  \bibfield  {author} {\bibinfo {author} {\bibfnamefont {C.~H.}\ \bibnamefont
  {Lee}},\ }\href@noop {} {\bibinfo {title} {Exceptional boundary states and
  negative entanglement entropy}} (\bibinfo {year} {2020}),\ \Eprint
  {https://arxiv.org/abs/2011.09505} {arXiv:2011.09505 [cond-mat.quant-gas]}
  \BibitemShut {NoStop}%
\bibitem [{\citenamefont {Wang}\ \emph {et~al.}(2015)\citenamefont {Wang},
  \citenamefont {Xu}, \citenamefont {Wang},\ and\ \citenamefont
  {Wu}}]{wa.xu.15}%
  \BibitemOpen
  \bibfield  {author} {\bibinfo {author} {\bibfnamefont {D.}~\bibnamefont
  {Wang}}, \bibinfo {author} {\bibfnamefont {S.}~\bibnamefont {Xu}}, \bibinfo
  {author} {\bibfnamefont {Y.}~\bibnamefont {Wang}},\ and\ \bibinfo {author}
  {\bibfnamefont {C.}~\bibnamefont {Wu}},\ }\href
  {https://doi.org/10.1103/PhysRevB.91.115118} {\bibfield  {journal} {\bibinfo
  {journal} {Phys. Rev. B}\ }\textbf {\bibinfo {volume} {91}},\ \bibinfo
  {pages} {115118} (\bibinfo {year} {2015})}\BibitemShut {NoStop}%
\bibitem [{\citenamefont {Li}\ \emph {et~al.}(2020)\citenamefont {Li},
  \citenamefont {Lee},\ and\ \citenamefont {Gong}}]{li.le.20}%
  \BibitemOpen
  \bibfield  {author} {\bibinfo {author} {\bibfnamefont {L.}~\bibnamefont
  {Li}}, \bibinfo {author} {\bibfnamefont {C.~H.}\ \bibnamefont {Lee}},\ and\
  \bibinfo {author} {\bibfnamefont {J.}~\bibnamefont {Gong}},\ }\href
  {https://doi.org/10.1103/PhysRevLett.124.250402} {\bibfield  {journal}
  {\bibinfo  {journal} {Phys. Rev. Lett.}\ }\textbf {\bibinfo {volume} {124}},\
  \bibinfo {pages} {250402} (\bibinfo {year} {2020})}\BibitemShut {NoStop}%
\bibitem [{\citenamefont {Brody}(2013)}]{brod.13}%
  \BibitemOpen
  \bibfield  {author} {\bibinfo {author} {\bibfnamefont {D.~C.}\ \bibnamefont
  {Brody}},\ }\href@noop {} {\bibfield  {journal} {\bibinfo  {journal} {Journal
  of Physics A: Mathematical and Theoretical}\ }\textbf {\bibinfo {volume}
  {47}},\ \bibinfo {pages} {035305} (\bibinfo {year} {2013})}\BibitemShut
  {NoStop}%
\bibitem [{\citenamefont {Peschel}(2003)}]{pesc.03}%
  \BibitemOpen
  \bibfield  {author} {\bibinfo {author} {\bibfnamefont {I.}~\bibnamefont
  {Peschel}},\ }\href {https://doi.org/10.1088/0305-4470/36/14/101} {\bibfield
  {journal} {\bibinfo  {journal} {Journal of Physics A: Mathematical and
  General}\ }\textbf {\bibinfo {volume} {36}},\ \bibinfo {pages} {L205}
  (\bibinfo {year} {2003})}\BibitemShut {NoStop}%
\bibitem [{\citenamefont {Cheong}\ and\ \citenamefont
  {Henley}(2004)}]{ch.he.04}%
  \BibitemOpen
  \bibfield  {author} {\bibinfo {author} {\bibfnamefont {S.-A.}\ \bibnamefont
  {Cheong}}\ and\ \bibinfo {author} {\bibfnamefont {C.~L.}\ \bibnamefont
  {Henley}},\ }\href {https://doi.org/10.1103/PhysRevB.69.075111} {\bibfield
  {journal} {\bibinfo  {journal} {Phys. Rev. B}\ }\textbf {\bibinfo {volume}
  {69}},\ \bibinfo {pages} {075111} (\bibinfo {year} {2004})}\BibitemShut
  {NoStop}%
\bibitem [{\citenamefont {Peschel}\ and\ \citenamefont
  {Eisler}(2009)}]{pe.ei.09}%
  \BibitemOpen
  \bibfield  {author} {\bibinfo {author} {\bibfnamefont {I.}~\bibnamefont
  {Peschel}}\ and\ \bibinfo {author} {\bibfnamefont {V.}~\bibnamefont
  {Eisler}},\ }\href {https://doi.org/10.1088/1751-8113/42/50/504003}
  {\bibfield  {journal} {\bibinfo  {journal} {Journal of Physics A:
  Mathematical and Theoretical}\ }\textbf {\bibinfo {volume} {42}},\ \bibinfo
  {pages} {504003} (\bibinfo {year} {2009})}\BibitemShut {NoStop}%
\bibitem [{\citenamefont {Bai}\ \emph {et~al.}(2000)\citenamefont {Bai},
  \citenamefont {Demmel}, \citenamefont {Dongarra}, \citenamefont {Ruhe},\ and\
  \citenamefont {van~der Vorst}}]{ba.de.00}%
  \BibitemOpen
  \bibinfo {editor} {\bibfnamefont {Z.}~\bibnamefont {Bai}}, \bibinfo {editor}
  {\bibfnamefont {J.}~\bibnamefont {Demmel}}, \bibinfo {editor} {\bibfnamefont
  {J.}~\bibnamefont {Dongarra}}, \bibinfo {editor} {\bibfnamefont
  {A.}~\bibnamefont {Ruhe}},\ and\ \bibinfo {editor} {\bibfnamefont
  {H.}~\bibnamefont {van~der Vorst}},\ eds.,\ \bibinfo {title} {Templates for
  the solution of algebraic eigenvalue problems: A practical guide}\ (\bibinfo
  {publisher} {SIAM, Philadelphia},\ \bibinfo {year} {2000})\BibitemShut
  {NoStop}%
\bibitem [{\citenamefont {Lin}(1990)}]{lin.90}%
  \BibitemOpen
  \bibfield  {author} {\bibinfo {author} {\bibfnamefont {H.~Q.}\ \bibnamefont
  {Lin}},\ }\href {https://doi.org/10.1103/PhysRevB.42.6561} {\bibfield
  {journal} {\bibinfo  {journal} {Phys. Rev. B}\ }\textbf {\bibinfo {volume}
  {42}},\ \bibinfo {pages} {6561} (\bibinfo {year} {1990})}\BibitemShut
  {NoStop}%
\bibitem [{\citenamefont {Lin}\ \emph {et~al.}(1993)\citenamefont {Lin},
  \citenamefont {Gubernatis}, \citenamefont {Gould},\ and\ \citenamefont
  {Tobochnik}}]{li.gu.93}%
  \BibitemOpen
  \bibfield  {author} {\bibinfo {author} {\bibfnamefont {H.}~\bibnamefont
  {Lin}}, \bibinfo {author} {\bibfnamefont {J.}~\bibnamefont {Gubernatis}},
  \bibinfo {author} {\bibfnamefont {H.}~\bibnamefont {Gould}},\ and\ \bibinfo
  {author} {\bibfnamefont {J.}~\bibnamefont {Tobochnik}},\ }\href@noop {}
  {\bibfield  {journal} {\bibinfo  {journal} {Computers in Physics}\ }\textbf
  {\bibinfo {volume} {7}},\ \bibinfo {pages} {400} (\bibinfo {year}
  {1993})}\BibitemShut {NoStop}%
\bibitem [{\citenamefont {Kawabata}\ \emph {et~al.}(2019)\citenamefont
  {Kawabata}, \citenamefont {Shiozaki}, \citenamefont {Ueda},\ and\
  \citenamefont {Sato}}]{ka.sh.19}%
  \BibitemOpen
  \bibfield  {author} {\bibinfo {author} {\bibfnamefont {K.}~\bibnamefont
  {Kawabata}}, \bibinfo {author} {\bibfnamefont {K.}~\bibnamefont {Shiozaki}},
  \bibinfo {author} {\bibfnamefont {M.}~\bibnamefont {Ueda}},\ and\ \bibinfo
  {author} {\bibfnamefont {M.}~\bibnamefont {Sato}},\ }\href
  {https://doi.org/10.1103/PhysRevX.9.041015} {\bibfield  {journal} {\bibinfo
  {journal} {Phys. Rev. X}\ }\textbf {\bibinfo {volume} {9}},\ \bibinfo {pages}
  {041015} (\bibinfo {year} {2019})}\BibitemShut {NoStop}%
\bibitem [{\citenamefont {Calabrese}\ and\ \citenamefont
  {Cardy}(2004)}]{ca.ca.04}%
  \BibitemOpen
  \bibfield  {author} {\bibinfo {author} {\bibfnamefont {P.}~\bibnamefont
  {Calabrese}}\ and\ \bibinfo {author} {\bibfnamefont {J.}~\bibnamefont
  {Cardy}},\ }\href {https://doi.org/10.1088/1742-5468/2004/06/p06002}
  {\bibfield  {journal} {\bibinfo  {journal} {Journal of Statistical Mechanics:
  Theory and Experiment}\ }\textbf {\bibinfo {volume} {2004}},\ \bibinfo
  {pages} {P06002} (\bibinfo {year} {2004})}\BibitemShut {NoStop}%
\bibitem [{\citenamefont {Kim}(2014)}]{kim.14}%
  \BibitemOpen
  \bibfield  {author} {\bibinfo {author} {\bibfnamefont {I.~H.}\ \bibnamefont
  {Kim}},\ }\href {https://doi.org/10.1103/PhysRevB.89.235120} {\bibfield
  {journal} {\bibinfo  {journal} {Phys. Rev. B}\ }\textbf {\bibinfo {volume}
  {89}},\ \bibinfo {pages} {235120} (\bibinfo {year} {2014})}\BibitemShut
  {NoStop}%
\bibitem [{\citenamefont {Fromholz}\ \emph {et~al.}(2020)\citenamefont
  {Fromholz}, \citenamefont {Magnifico}, \citenamefont {Vitale}, \citenamefont
  {Mendes-Santos},\ and\ \citenamefont {Dalmonte}}]{fr.ma.20}%
  \BibitemOpen
  \bibfield  {author} {\bibinfo {author} {\bibfnamefont {P.}~\bibnamefont
  {Fromholz}}, \bibinfo {author} {\bibfnamefont {G.}~\bibnamefont {Magnifico}},
  \bibinfo {author} {\bibfnamefont {V.}~\bibnamefont {Vitale}}, \bibinfo
  {author} {\bibfnamefont {T.}~\bibnamefont {Mendes-Santos}},\ and\ \bibinfo
  {author} {\bibfnamefont {M.}~\bibnamefont {Dalmonte}},\ }\href
  {https://doi.org/10.1103/PhysRevB.101.085136} {\bibfield  {journal} {\bibinfo
   {journal} {Phys. Rev. B}\ }\textbf {\bibinfo {volume} {101}},\ \bibinfo
  {pages} {085136} (\bibinfo {year} {2020})}\BibitemShut {NoStop}%
\bibitem [{\citenamefont {Ye}\ \emph {et~al.}(2016)\citenamefont {Ye},
  \citenamefont {Mu},\ and\ \citenamefont {Fan}}]{ye.mu.16}%
  \BibitemOpen
  \bibfield  {author} {\bibinfo {author} {\bibfnamefont {B.-T.}\ \bibnamefont
  {Ye}}, \bibinfo {author} {\bibfnamefont {L.-Z.}\ \bibnamefont {Mu}},\ and\
  \bibinfo {author} {\bibfnamefont {H.}~\bibnamefont {Fan}},\ }\href
  {https://doi.org/10.1103/PhysRevB.94.165167} {\bibfield  {journal} {\bibinfo
  {journal} {Phys. Rev. B}\ }\textbf {\bibinfo {volume} {94}},\ \bibinfo
  {pages} {165167} (\bibinfo {year} {2016})}\BibitemShut {NoStop}%
\bibitem [{\citenamefont {Kozii}\ and\ \citenamefont {Fu}(2017)}]{ko.fu.17}%
  \BibitemOpen
  \bibfield  {author} {\bibinfo {author} {\bibfnamefont {V.}~\bibnamefont
  {Kozii}}\ and\ \bibinfo {author} {\bibfnamefont {L.}~\bibnamefont {Fu}},\
  }\href@noop {} {\bibinfo {title} {Non-hermitian topological theory of
  finite-lifetime quasiparticles: Prediction of bulk fermi arc due to
  exceptional point}} (\bibinfo {year} {2017}),\ \Eprint
  {https://arxiv.org/abs/1708.05841} {arXiv:1708.05841 [cond-mat.mes-hall]}
  \BibitemShut {NoStop}%
\bibitem [{\citenamefont {Nagai}\ \emph {et~al.}(2020)\citenamefont {Nagai},
  \citenamefont {Qi}, \citenamefont {Isobe}, \citenamefont {Kozii},\ and\
  \citenamefont {Fu}}]{na.qi.20}%
  \BibitemOpen
  \bibfield  {author} {\bibinfo {author} {\bibfnamefont {Y.}~\bibnamefont
  {Nagai}}, \bibinfo {author} {\bibfnamefont {Y.}~\bibnamefont {Qi}}, \bibinfo
  {author} {\bibfnamefont {H.}~\bibnamefont {Isobe}}, \bibinfo {author}
  {\bibfnamefont {V.}~\bibnamefont {Kozii}},\ and\ \bibinfo {author}
  {\bibfnamefont {L.}~\bibnamefont {Fu}},\ }\href
  {https://doi.org/10.1103/PhysRevLett.125.227204} {\bibfield  {journal}
  {\bibinfo  {journal} {Phys. Rev. Lett.}\ }\textbf {\bibinfo {volume} {125}},\
  \bibinfo {pages} {227204} (\bibinfo {year} {2020})}\BibitemShut {NoStop}%
\bibitem [{\citenamefont {Crippa}\ \emph {et~al.}(2021)\citenamefont {Crippa},
  \citenamefont {Budich},\ and\ \citenamefont {Sangiovanni}}]{cr.bu.21}%
  \BibitemOpen
  \bibfield  {author} {\bibinfo {author} {\bibfnamefont {L.}~\bibnamefont
  {Crippa}}, \bibinfo {author} {\bibfnamefont {J.~C.}\ \bibnamefont {Budich}},\
  and\ \bibinfo {author} {\bibfnamefont {G.}~\bibnamefont {Sangiovanni}},\
  }\href {https://doi.org/10.1103/PhysRevB.104.L121109} {\bibfield  {journal}
  {\bibinfo  {journal} {Phys. Rev. B}\ }\textbf {\bibinfo {volume} {104}},\
  \bibinfo {pages} {L121109} (\bibinfo {year} {2021})}\BibitemShut {NoStop}%
\bibitem [{\citenamefont {Mitscherling}\ and\ \citenamefont
  {Metzner}(2021)}]{mi.me.21}%
  \BibitemOpen
  \bibfield  {author} {\bibinfo {author} {\bibfnamefont {J.}~\bibnamefont
  {Mitscherling}}\ and\ \bibinfo {author} {\bibfnamefont {W.}~\bibnamefont
  {Metzner}},\ }\href {https://doi.org/10.1103/PhysRevB.104.L201107} {\bibfield
   {journal} {\bibinfo  {journal} {Phys. Rev. B}\ }\textbf {\bibinfo {volume}
  {104}},\ \bibinfo {pages} {L201107} (\bibinfo {year} {2021})}\BibitemShut
  {NoStop}%
\end{thebibliography}


\end{document}